\documentclass[iop]{emulateapj}
\usepackage{apjfonts}
\usepackage{epsfig,color}
\usepackage{mathrsfs}
\usepackage{ulem}
\usepackage{amsmath}
\usepackage{empheq}
\usepackage{bm}
\usepackage{longtable}
\usepackage{rotating}
\usepackage{natbib}
\usepackage{subfigure}
\usepackage{booktabs}

\makeatletter

\newcommand{\Rmnum}[1]{\expandafter\@slowromancap\romannumeral #1@}
\makeatother
\newcommand{\xmm}{\hbox{\it XMM-Newton\/}}
\newcommand{\chandra}{{\it Chandra\/}}
\newcommand{\xray}{\hbox{\it \rm X-ray\/}}
\newcommand{\erosita}{\hbox{\it eROSITA\/}}
\newcommand{\edd}{$\lambda_{\rm Edd}$}
\newcommand{\mdot}{$\dot{\mathscr{M}}$}
\newcommand{\mdotnew}{$\dot{\mathscr{M}}$}
\newcommand{\mdotold}{$\dot{\mathscr{M}}_{NT}$}
\newcommand{\hb}{${\rm H\beta}$}
\newcommand{\gam}{${\Gamma}$}

\newcommand{\mbh}{${{M}_{\rm BH}}$}
\newcommand{\mbhtd}{${{M}_{\rm BH}}$}
\newcommand{\tnmbh}{${{M}_{\rm BH,NT}}$}
\newcommand{\ledd}{${{L}_{\rm Edd}}$}
\newcommand{\lbol}{${{L}_{\rm Bol}}$}
\newcommand{\lbolbc}{${{L}_{\rm Bol,BC}}$}
\newcommand{\lbolsed}{${{L}_{\rm Bol}}$}
\newcommand{\newedd}{$\lambda_{\rm Edd}$}
\newcommand{\tnedd}{$\lambda_{\rm Edd,NT}$}

\begin{document}
\title{On the relation between hard X-ray photon index versus Accretion Rate for \hbox{super-Eddington} accreting quasars}

\author{
Jian~Huang\altaffilmark{1,2,3},
Bin~Luo\altaffilmark{1,2,3},
Pu~Du\altaffilmark{4},
Chen~Hu\altaffilmark{4},
Jian-Min~Wang\altaffilmark{4,5,6},
and Yi-Jia~Li\altaffilmark{7}
}

\altaffiltext{1}{School of Astronomy and Space Science, Nanjing University, Nanjing, Jiangsu 210093, China; bluo@nju.edu.cn}
\altaffiltext{2}{Key Laboratory of Modern Astronomy and Astrophysics (Nanjing University), Ministry of Education, Nanjing 210093, China}
\altaffiltext{3}{Collaborative Innovation Center of Modern Astronomy and Space Exploration, Nanjing 210093, China}
\altaffiltext{4}{Key Laboratory for Particle Astrophysics, Institute of High Energy Physics, Chinese Academy of Sciences, 19B Yuquan Road, Beijing 100049, China}
\altaffiltext{5}{School of Astronomy and Space Science, University of Chinese Academy of Sciences, 19A Yuquan Road, Beijing 100049, China}
\altaffiltext{6}{National Astronomical Observatories of China, Chinese Academy of Sciences, 20A Datun Road, Beijing 100020, China}
\altaffiltext{7}{Department of Astronomy \& Astrophysics, 525 Davey Lab,
The Pennsylvania State University, University Park, PA 16802, USA}

\begin{abstract}
We investigate whether the hard X-ray photon index (\gam) versus accretion rate correlation for \hbox{super-Eddington} accreting quasars is different from that for \hbox{sub-Eddington} accreting quasars.
We construct a sample of 113 bright quasars from the Sloan Digital Sky Survey Data Release 14 quasar catalog, including 38 quasars as the \hbox{super-Eddington} subsample and 75 quasars as the \hbox{sub-Eddington} subsample.
We derive black-hole masses using a simple-epoch virial mass formula based on the \hb\ lines, and we use the standard thin disk model to derive the dimensionless accretion rates (\mdot) for our sample.
The \xray\ data for these quasars are collected from the \chandra\ and \xmm\ archives.
We fit the hard \xray\ spectra using a single \hbox{power-law} model to obtain \gam\ values.
We find a statistically significant ($R_{\rm S}=0.43$, $p=7.75\times{10}^{-3}$) correlation between \gam\ and \mdot\ for the \hbox{super-Eddington} subsample.
The \hbox{\gam--\mdot} correlation for the \hbox{sub-Eddington} subsample is also significant, but weaker ($R_{\rm S}=0.30$, $p=9.98\times{10}^{-3}$).
Linear regression analysis shows that \hbox{${\rm \Gamma}=(0.34\pm0.11){\rm log}{\dot{\mathscr{M}}}+(1.71\pm0.17)$} and \hbox{${\rm 
\Gamma}=(0.09\pm0.04){\rm log}{\dot{\mathscr{M}}}+(1.93\pm0.04)$} for the super- and \hbox{sub-Eddington} subsamples, respectively.
The \hbox{\gam--\mdot} correlations of the two subsamples are different, suggesting different disk--corona connections in these two types of systems.
We propose one qualitative explanation of the steeper \hbox{\gam--\mdot} correlation in the \hbox{super-Eddington} regime that involves larger seed photon fluxes received by the compact coronae from the thick disks in \hbox{super-Eddington} accreting quasars.

\end{abstract}
\keywords{galaxies: active  -- quasars: general -- X-rays: galaxies}

\section{Introduction}\label{sec:intro}

Active Galactic Nuclei (AGNs) produce considerable amount of \xray\ emission ubiquitously \citep{Tananbaum79}.
It is considered to be produced by a corona of hot electrons located close to the inner accretion disk of the supermassive \hbox{black-hole} (SMBH).
The optical/UV photons from the thermal accretion disk emission are inverse-Compton scattered into the \xray\ energies by the hot electrons in the corona \citep[e.g.,][]{Liang1977,Francesco1993,Done2010,Gilfanov2014,Fabian2017}.
Such a mechanism implies a connection between the \xray\ coronae and the accretion disks for AGNs.

Previous studies have found a significant positive correlation between the hard (rest-frame $>2$ keV) \xray\ photon index (\gam)\footnote{The hard \xray\ photon spectra of AGNs are usually described by a \hbox{power-law} form, $N_{E}\propto{E}^{-\Gamma}$(photons ${\rm cm}^{-2}~{\rm s}^{-1}~{\rm keV}^{-1}$).} and the accretion rate parameterized as the Eddington ratio (\edd) for typical AGNs \citep[e.g.,][]{lu1999,wang2004,Shemmer2008, Risaliti2009, Brightman2013, Trakhtenbrot2017}, where \edd~=~\lbol$/$\ledd\ with \lbol\ being the bolometric luminosity and \ledd\ the Eddington luminosity.
This \hbox{\gam--\edd} correlation is indicative of the connection between the accretion disk and \xray\ corona.
The physics behind this correlation is not clear.
One possible explanation is that when the accretion rate is higher, the cooling of the corona becomes more efficient, which decreases the temperature and/or the optical depth of the corona \citep[e.g.,][]{Fabian2015,Kara2017,Ricci2018,Barua2020}.
The \xray\ spectrum is thus softer because the cooler corona produces relatively fewer hard \xray\ photons \citep[e.g.,][]{Vasudevan2007,Davis2011}.

Previous studies typically found substantial scatter for the \hbox{\gam--\edd} correlation, which may be partially due to the complications in deriving the \edd\ and \gam\ parameters.
For example, there are substantial uncertainties for the \hbox{single-epoch} virial \hbox{black-hole} masses, including systematic errors ($\approx$\hbox{0.4--0.5 dex}; e.g., \citealt{Shen2013}) and measurement errors ($\approx$0.15 dex; \citealt{Shen2011}).
There may be even bias when using the \ion{C}{4} line emission to estimate \hbox{black-hole} masses, due to the possible blueshift component of \ion{C}{4} which may be produced by outflowing wind \citep[e.g.,][]{Baskin2005,Richards2011,Shen2013,plotkin2015}.
The different methods and energy bands used for fitting the \xray\ spectra may in addition lead to systematically different \gam\ values \citep[e.g.,][]{Trakhtenbrot2017,Ricci2018}.

The different physics of accretion disks with different accretion rates may also affect the observed \hbox{\gam--\edd} correlation.
The accretion disk of \hbox{sub-Eddington} accreting AGNs with normal accretion rates ($0.001\lesssim$~\edd~$\lesssim0.1$; e.g., \citealt{Netzer2019}) is generally described as a geometrically thin, optically thick accretion disk \citep{Shakura73}.
For \hbox{super-Eddington} accreting AGNs with high accretion rates (\edd~$\gtrsim{0.1}$), geometrically thick inner accretion disks are generally expected from either analytical solutions \citep[e.g.,][]{Abramowicz1988,wang1999,Mineshige2000} or numerical simulations \citep[e.g.,][]{Sadowski2016,Jiang2014,Jiang2019}.
In analytical solutions, the geometrically thick accretion disk in the \hbox{super-Eddington} regime has the `photon trapping' effect \citep[e.g.,][]{Abramowicz1988,wang1999,ohsuga2002}.
The diffuse timescale for photons to escape from the thick disk surface may be longer than the timescale for photons to be advected into the central black hole.
Therefore, the bolometric luminosity of a \hbox{super-Eddington} accreting AGN may be saturated and depend weakly on its accretion rate, which can be expressed as \hbox{${L_{\rm Bol}}\approx 2{L_{\rm Edd}}[1+{\rm ln}(\dot{\mathscr{M}}/50)]$} \citep[e.g.,][]{Mineshige2000,wang2014a}.
In the equation above, the dimensionless accretion rate (\mdot) is defined as \hbox{\mdot=$\dot{M}c^2/L_{\rm Edd}$}, where $\dot{M}$ is the mass accretion rate; \mdot\ is related to \edd\ as \mdot$=$\edd$/\eta$, where $\eta$ is the radiative efficiency parameter.
However, recent simulation results suggest that photons may escape from the thick disk surface more efficiently via vertical advection from effects such as magnetic buoyancy \citep[e.g.,][]{Jiang2014,Sadowski2014}.
These suggest that the photon trapping effect may not dominate the cooling of accretion disks in the \hbox{super-Eddington} regime. 

The differences of the disk structure and physics between sub- and \hbox{super-Eddington} accreting AGNs suggest that the connections between the accretion disk and the \xray\ corona may be different in these two types of systems.
Observationally, we may expect different correlations between the \xray\ photon index and Eddington ratio, which contribute partially to the strong scatter of the overall \hbox{\gam--\edd} correlation.
Previous studies on the \hbox{\gam--\edd} correlation did not separate these two populations of AGNs into respective samples, probably because of the limited sample sizes and the difficulty in selecting \hbox{super-Eddington} accreting AGNs.
For \hbox{super-Eddington} accreting AGNs, the Eddington ratio may not be a good indicator of the accretion rate due to the possible photon trapping effect.
Therefore, it is valuable to check the \hbox{\gam--\mdot} correlations when investigating the connections between the accretion disks and coronae for \hbox{super-Eddington} accreting AGNs.

The \hbox{black-hole} mass is the key parameter for computing the dimensionless accretion rate or the Eddington ratio.
For distant AGNs, the \hbox{black-hole} masses are usually estimated using \hbox{single-epoch} virial mass estimation that is based on the empirical broad line region (BLR) size versus luminosity (\hbox{$R$--${L}$}) relation \citep[e.g.,][]{Kaspi2000,Kaspi2005,Netzer2007}.
Recently, it has been proposed that the conventional \hbox{$R$--${L}$} relation may overestimate the BLR sizes for \hbox{super-Eddington} accreting AGNs \citep[e.g.,][]{wang2014a,du2016}, and the virial \hbox{black-hole} masses are thus overestimated.
Based on the analysis of a sample of AGNs with reverberation mapping data including a sample of \hbox{super-Eddington} accreting AGNs, \cite{du2019} take into account the \ion{Fe}{2} emission strength (${\cal{R}_{\rm Fe}}$)\footnote{ ${\cal{R}_{\rm Fe}}$ is the relative strength of the \ion{Fe}{2} line emission in the \hbox{rest-frame} \hbox{4434--4684} \AA\ band to the broad H${\beta}$ emission line (${L_{\rm Fe}}/{L_{\rm H\beta}}$, where ${L_{\rm Fe}}$ and ${L_{\rm H\beta}}$ are the luminosities of the \ion{Fe}{2} and broad H${\beta}$ emission line, respectively).} and propose an updated \hbox{$R$--${L}$} relation to provide more accurate estimations of \hbox{black-hole} masses, especially for \hbox{super-Eddington} accreting AGNs.

In this study, we aim to investigate if there is any difference between the disk--corona connections in super- and \hbox{sub-Eddington} accreting AGNs, by comparing the correlations between the hard \xray\ photon index and the accretion rates for these two types of systems.
Due to the additional uncertainties on the estimations of bolometric luminosities, especially in the \hbox{super-Eddington} regime, we prioritize our investigation in the \hbox{\gam--\mdot} correlation.
Statistically significant samples of super- and \hbox{sub-Eddington} accreting AGNs are thus needed.
The Slon Digital Sky Survey (SDSS) Data Release 14 (DR14) quasar catalog \citep{Paris2018} provides a large sample of quasars with optical spectra.
The \xray\ data of these SDSS quasars can be searched from the \chandra\ and \xmm\ archives.
We use broad ${\rm H\beta}$ \hbox{emission-line profiles} and the updated \hbox{$R$--${L}$} relation of \cite{du2019} for relatively reliable \hbox{black-hole} mass estimation and \hbox{super-Eddington} accreting quasar selection.

We organize our work as follows.
In Section~\ref{sec:s} we present our sample selection using the SDSS DR14 quasar catalog and the \chandra\ and \xmm\ archives.
Basic quasar properties including \hbox{black-hole} masses, dimensionless accretion rates and Eddington ratios are derived for our final sample.
In Section~\ref{sec:data} we describe the procedure for \xray\ data reduction, and we measure the \gam\ values for our final sample.
In Section~\ref{sec:results} we investigate the correlation between the hard \xray\ photon index and dimensionless accretion rates for the \hbox{super-Eddington} subsample, and we compare it to that of the \hbox{sub-Eddington} subsample. 
In Section~\ref{sec:dis} we discuss the implication of our results.
In Section~\ref{sec:con} we summary our work and discuss some future prospects.

Throughout this paper, we use a cosmology with $H_0=67.4$~km~s$^{-1}$~Mpc$^{-1}$, $\Omega_{\rm M}=0.315$, and $\Omega_{\Lambda}=0.685$ \citep{Planck2018}.

\section{Sample Selection}\label{sec:s}
\subsection{Initial SDSS Quasar Selection} \label{sec:initialsample}
We use the SDSS DR14 quasar catalog \citep{Paris2018}, which contains 526\,356 quasars, to select an initial quasar sample.
We first select 36\,697 quasars with ${z < 0.7}$.
Within this redshift range, the SDSS spectra cover the \hbox{rest-frame} 5100~\AA\ continuum, the optical \ion{Fe}{2} line emission, and the broad \hb\ emission line, so that we can measure the \hbox{rest-frame} \hbox{5100~\AA} continuum luminosities, the ${\cal{R}_{\rm Fe}}$ values, and the full width at half maximum (FWHM) of the broad \hb\ emission line.
These parameters are used to derive the bolometric luminosities and \hbox{black-hole} masses.

Then we select bright quasars by requiring the $i$-band magnitude ($m_{i}$) to be less than 19, because the probability of finding useful \xray\ archival data for bright quasars is relatively high.
There are 12\,638 quasars satisfying both the redshift and $m_{i}$ criteria.
Before fitting the SDSS spectra of these quasars, we search for \xray\ archival coverage to reduce significantly the sample size and our workload.

\subsection{\chandra\ Archival Coverage} \label{sec:chandra}

We search for public \chandra\ Advanced CCD Imaging Spectrometer (ACIS) non-grating observations of all the 12\,638 initial sample objects in the \chandra\ archive\footnote{https://cda.harvard.edu/chaser/.} as of 2019 July 9.
For each quasar, we use a 14\arcmin\ matching radius to search for available \xray\ observations.
We find 903 quasars that have matched \chandra\ observations, including 205 quasars with multiple \chandra\ observations.

To select observations which yield large numbers of source counts for spectral fitting, we further filter the 903 quasars using the following criteria.

(1) The quasar is the target of the matched observation with an off-axis angle smaller than 1\arcmin.
We obtain 163 quasars after using this criterion.

(2) The quasar is not the target of the matched observation, but it has an off-axis angle smaller than 10\arcmin\ and the exposure time of the observation is longer than 5 ks.
We obtain 305 quasars using this criterion.

(3) The quasar has an \hbox{off-axis} angle larger than 10\arcmin\ but smaller than 14\arcmin, and the exposure time of the matched observation is longer than 20 ks.
We obtain 83 quasars using this criterion.

If a quasar still has multiple \chandra\ observations after the above selection, we only use the observation with the longest exposure time.
Using the criteria above, We select 551 (163+305+83) quasars with good archival ACIS data.
We analyze the \chandra\ data of these quasars to obtain their \xray\ properties (see Section~\ref{sec:cdata} below).
In order to obtain reliable spectral fitting results, we select only 120 of these 551 quasars with numbers of net counts more than 200 in the \hbox{observed-frame} \hbox{$2/(1+z)$--7~keV} band, excluding the Fe K complex that is adopted to be between \hbox{rest-frame} 5.5~keV and 7.5~keV \citep[e.g.,][]{Brightman2013}.

\subsection{\xmm\ Archival Coverage} \label{sec:xmm}

We use the \hbox{3XMM-DR8} source catalog\footnote{https://heasarc.gsfc.nasa.gov/w3browse/xmm-newton/xmmssc.html.} \citep{Watson2009,Rosen2016}, which contains 775\,153 \xray\ sources drawn from 10\,242 European Photon Imaging Camera (EPIC) observations between 2000 February 3 and 2017 November 30, to search for \xmm\ observations for the initial sample.
We match the 12\,638 quasars to the \hbox{3XMM-DR8} source catalog using a 3\arcsec\ matching radius, and we obtain 487 matches.

Among these quasars, we further select 188 quasars which have more than 1100 total PN camera counts in the \hbox{observed-frame} \hbox{0.2--12~keV} band adopted from the \hbox{3XMM-DR8} source catalog. 
This source count criterion is chosen to yield $\gtrsim160$ net source counts in the \hbox{observed-frame} \hbox{$2/(1+z)$--10~keV} band excluding the Fe K complex, adopting a single \hbox{power-law} spectrum with \gam$=1.9$, $z=0.4$ (the mean redshift of the 487 quasars), and typical PN response files.
For a quasar with multiple \xmm\ observations, we select the observation with the highest number of source counts in the \hbox{observed-frame} \hbox{0.2--12~keV} band from the 3XMM-DR8 source catalog.

We analyze the corresponding \xmm\ data to obtain \xray\ properties of these 188 quasars (see Section~\ref{sec:xdata} below).
We keep 118 of these quasars with more than 200 net source counts in the \hbox{observed-frame} $2/(1+z)$--10~keV band excluding the Fe K complex to obtain reliable spectral fitting results.

\subsection{Selection by SDSS Spectral Quality} \label{sec:sdss}
Among the 120 quasars with \chandra\ observations and 118 quasars with \xmm\ observations, there are 26 quasars in common.
For each of these 26 quasars, we choose the \chandra\ or \xmm\ observation with a larger number of net source counts.
Thus we obtain a sample of 212 ($120+118-26$) quasars.
We fit the SDSS spectra of these quasars following the same procedure described in \cite{hu2008, hu2015}.
The sample is further filtered with the following additional criteria based on the spectral quality and shapes.

(1) We require signal-to-noise ratio (S/N) per pixel to be \hbox{$>10$} in the \hbox{rest-frame} \hbox{4430--5550~\AA} spectrum.
This wavelength range covers the ${\rm H\beta},~[{\rm O~\Rmnum{3}}]$, and optical \ion{Fe}{2} line emission.
After applying this criterion, we select 179 quasars.

(2) We require the \hbox{power-law} spectral slope of the decomposed optical continuum ($\alpha_{\lambda}$) to be $<0$.
This criterion is to exclude quasars with SDSS spectra that may have strong host galaxy contamination or be affected by heavy absorption.
We select 161 quasars after using this criterion.
The \hbox{emission-line} and continuum properties for our final sample are listed in Table~\ref{tbl-optical}.

\begin{figure}
\centerline{
\includegraphics[scale=0.50]{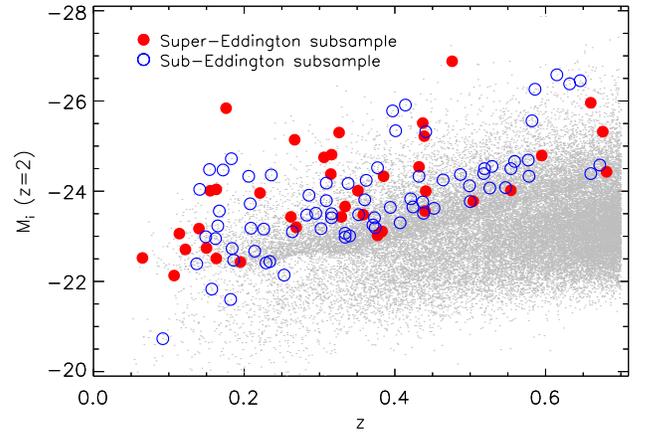}
}
\caption{Distribution of the absolute $i$-band magnitude (${M}_{\rm i}$) vs. redshift for our final sample. The ${M}_{\rm i}$ values are adopted from the SDSS DR14 quasar catalog. The red filled circles and blue open circles represent our super- and \hbox{sub-Eddington} subsample, respectively. The grey dots represent the SDSS DR14 quasars with ${z}<0.7$.  
}
\label{fig-magi}
\end{figure}

\subsection{Exclusion of Radio-Loud Quasars} \label{sec:rl}

Since radio-loud quasars may produce a significant amount of \xray\ emission associated with jets \citep[e.g.,][]{Miller2011}, we need to remove radio-loud quasars from our sample.
Following \cite{Shen2011}, we first match our sample of 161 quasars to the 14Dec17 version of the Faint Images of the Radio Sky at Twenty-Centimeters (FIRST) source catalog \citep{White1997} using a 3\arcsec\ matching radius.
For a quasar with two or more FIRST counterparts within the \hbox{30\arcsec-radius} circular region, we use the summed peak flux densities at \hbox{20 cm} of all the FIRST counterparts to compute the \hbox{rest-frame} \hbox{6 cm} flux density, $f_{\rm 6~cm}$, adopting a \hbox{power-law} spectral slope of ${\alpha_{\rm r}}=-0.8$ \citep[e.g.,][]{Falcke1996, Barvainis2005}.
There are 20 such quasars; we visually inspect the FIRST and Digital Sky Survey images of these sources, and we find no apparent additional optical counterparts associated with the FIRST counterparts, suggesting that these FIRST counterparts are radio components of the quasars.
For a quasar with only one FIRST counterpart within the matching radius, we rematch it to the FIRST catalog using a 5\arcsec\ matching radius to determine if the one FIRST source is the correct radio counterpart.
We then use the peak flux density at \hbox{20 cm} of the FIRST counterpart to compute the \hbox{rest-frame} \hbox{6 cm} flux density.
For a quasar with no FIRST counterpart, we set \hbox{5${\sigma}_{\rm rms}+0.25$ mJy} as the upper limit on the \hbox{20 cm} flux density, where ${\sigma}_{\rm rms}$ is the rms noise at the source position and \hbox{0.25 mJy} is used to account for the CLEAN bias \citep{Gibson2009}.
The upper limit on $f_{\rm 6~cm}$ is then calculated from the upper limit on the \hbox{20 cm} flux density.

There are six quasars not in the coverage of the FIRST catalog.
We match the six quasars to the NRAO VLA Sky Survey (NVSS) source catalog \citep{Condon1998} using the same method described above.
Only one quasar has a radio counterpart within a 5\arcsec\ matching radius.
For the other five quasars with no NVSS counterparts, we use \hbox{2.5 mJy} (the threshold of NVSS source detection) as the upper limits on the flux densities at \hbox{20 cm}.
We calculate $f_{\rm 6~cm}$ or its upper limit using the same method as that for quasars with FIRST coverage.

The sample of 161 quasars contains 56 quasars with FIRST or NVSS counterparts and 105 quasars without radio counterparts.
We convert the flux density at \hbox{rest-frame} 5100~{\AA} measured from the SDSS spectra (see Section~\ref{sec:sdss}) to the flux density at \hbox{rest-frame} 4400~{\AA} ($f_{\rm 4400}$) for each quasar using the optical \hbox{power-law} spectral slope we obtain from the spectral fitting.
We compute the radio-loudness parameter or its upper limit using \hbox{$R=f_{\rm 6~cm}/f_{\rm 4400~\AA}$} \citep[e.g.,][]{Kellermann1989}.
We consider a quasar to be radio quiet (RQ) if its $R$ value is less than 10 or its upper limit on $R$ is less than 100.
We remove 37 quasars with $R$ values more than 10 from our sample, and there is no quasar that has an upper limit on $R$ more than 100.
The remaining 124 quasars are considered to be RQ.
Only three quasars in these 124 quasars have upper limits on $R$ larger than 10 and the largest upper limit is only 13.6, suggesting that our sample is a reliable RQ quasar sample.
The radio properties for our final sample (see Section~\ref{sec:bal}) are listed in Table~\ref{tbl-optical}.

\subsection{Exclusion of X-ray Absorbed Quasars} \label{sec:bal}
Since we are studying the correlation between the corona and accretion disk, we need to obtain the intrinsic hard \xray\ photon index for our quasars.
The \xray\ emission from a small fraction of quasars may be affected by absorption, and the main population of these quasars are broad absorption line (BAL) quasars \citep[e.g.,][]{Gallagher2002,Gallagher2006,Fan2009,Gibson2009}.
It is difficult to derive the intrinsic \gam\ values for \xray\ absorbed quasars without very good \xray\ spectra.
Therefore, we need to remove \xray\ absorbed quasars from our sample.

We remove 11 \xray\ absorbed quasars which are probably BAL quasars from our sample after fitting the \xray\ spectra (see Section~\ref{sec:dataanalysis} below).
We check the SDSS spectra of these 11 quasars for BAL features.
Five quasars have no \ion{Mg}{2} coverage, and the other six quasars do not have apparent \ion{Mg}{2} absorption.
\ion{Mg}{2} BAL quasars are much rarer than \ion{C}{4} BAL quasars, but for the redshift range of our sample, the SDSS DR14 spectra do not cover the \ion{C}{4} line.
We note that the fraction of \xray\ absorbed quasars in this sample ($11/124$) is smaller than the fraction of BAL quasars (${\approx}$15$\%$; e.g., \citealt{Hewett2003,Trump2006,Gibson2009,Allen2011}).
We consider that this is a natural consequence of us selecting \xray\ bright quasars (see Section~\ref{sec:chandra} and Section~\ref{sec:xmm}), which generally guards against \xray\ absorbed quasars.

After excluding the 11 \xray\ absorbed quasars, the remaining 113 quasars constitute our final sample.
We show the distribution of the absolute $i$-band magnitude (at $z=2$; \citealt{Richards2006}) versus redshifts for the 113 quasars in Figure~\ref{fig-magi}.
Compared to typical $z<0.7$ SDSS DR14 quasars, these 113 quasars are at the luminous end of the absolute $i$-band magnitude distribution.

\begin{figure*}
\centerline{
\includegraphics[scale=0.50]{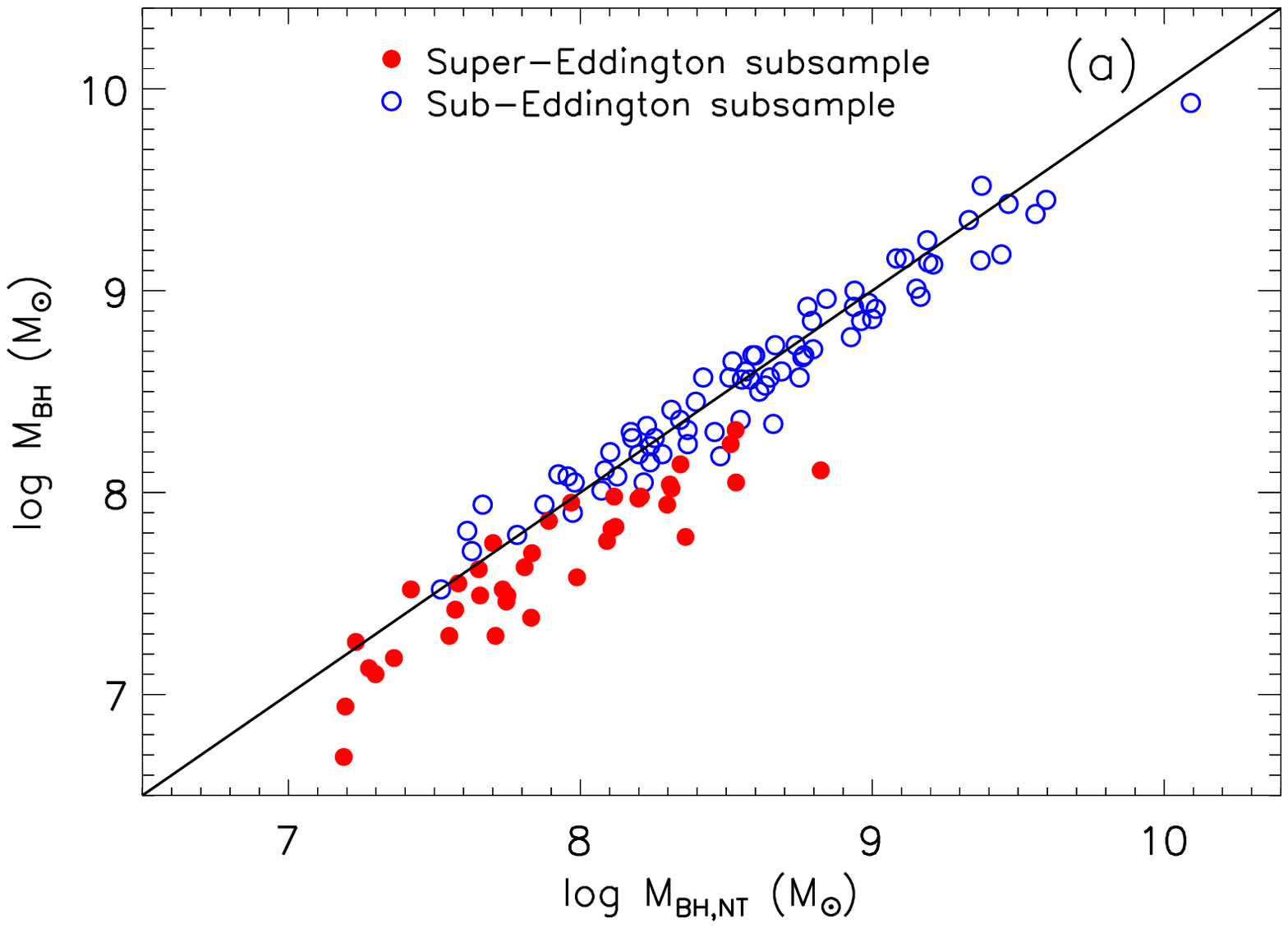}
\includegraphics[scale=0.50]{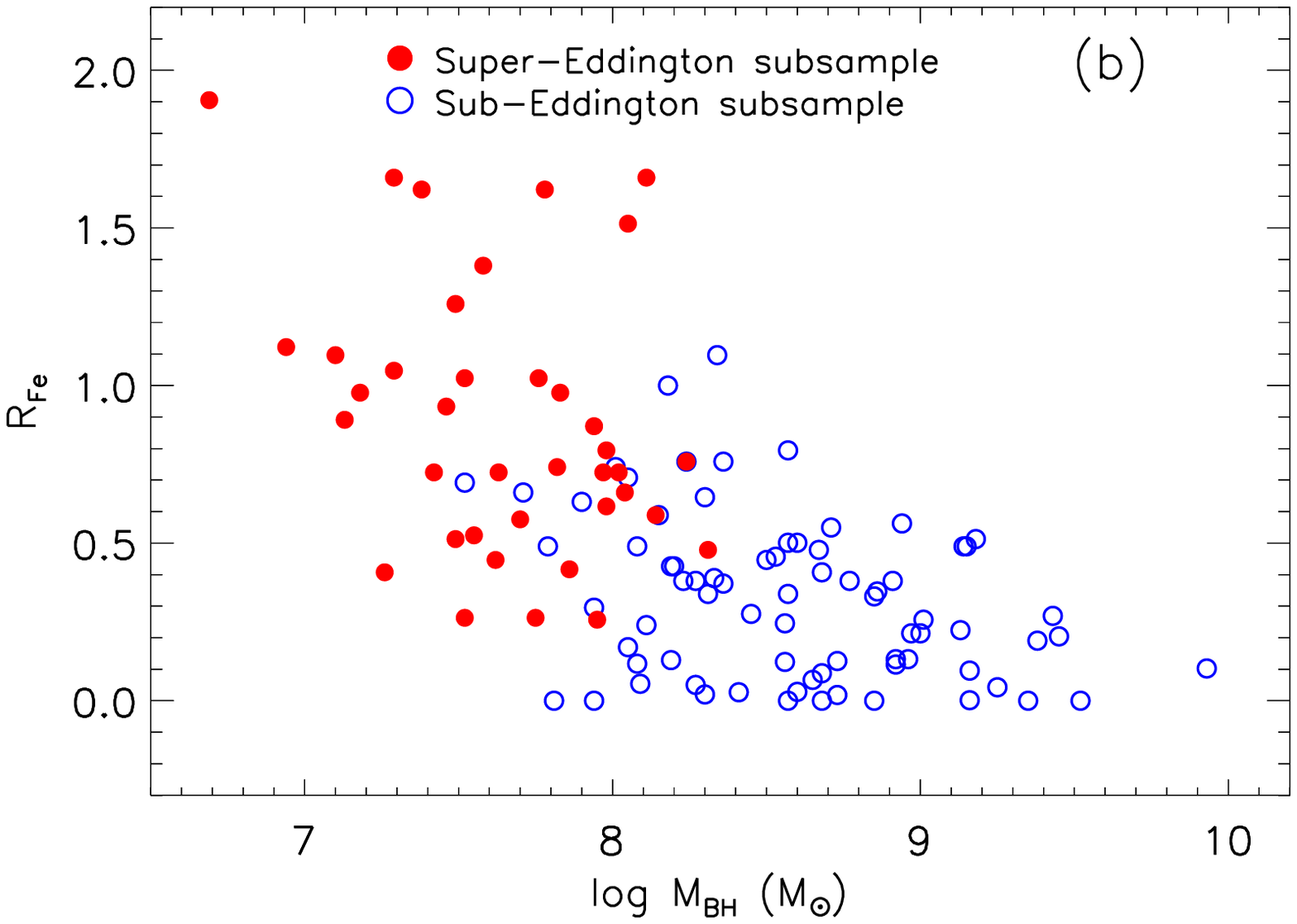}
}

\caption{
 {\bf (a):} \mbhtd\ vs. \tnmbh\ for the super- and \hbox{sub-Eddington} subsamples.
We derive the \mbhtd\ and \tnmbh\ values using Equation~\ref{2} and Equation~\ref{3}, respectively. 
 {\bf (b):} ${\cal{R}_{\rm Fe}}$ vs. \mbhtd\ for the super- and the \hbox{sub-Eddington} subsample.
The \hbox{super-Eddington} subsample has obviously higher ${\cal{R}_{\rm Fe}}$ values than those of the \hbox{sub-Eddington} subsample. 
}
\label{fig-mass}
\end{figure*}

\begin{figure*}
\centerline{
\includegraphics[scale=0.50]{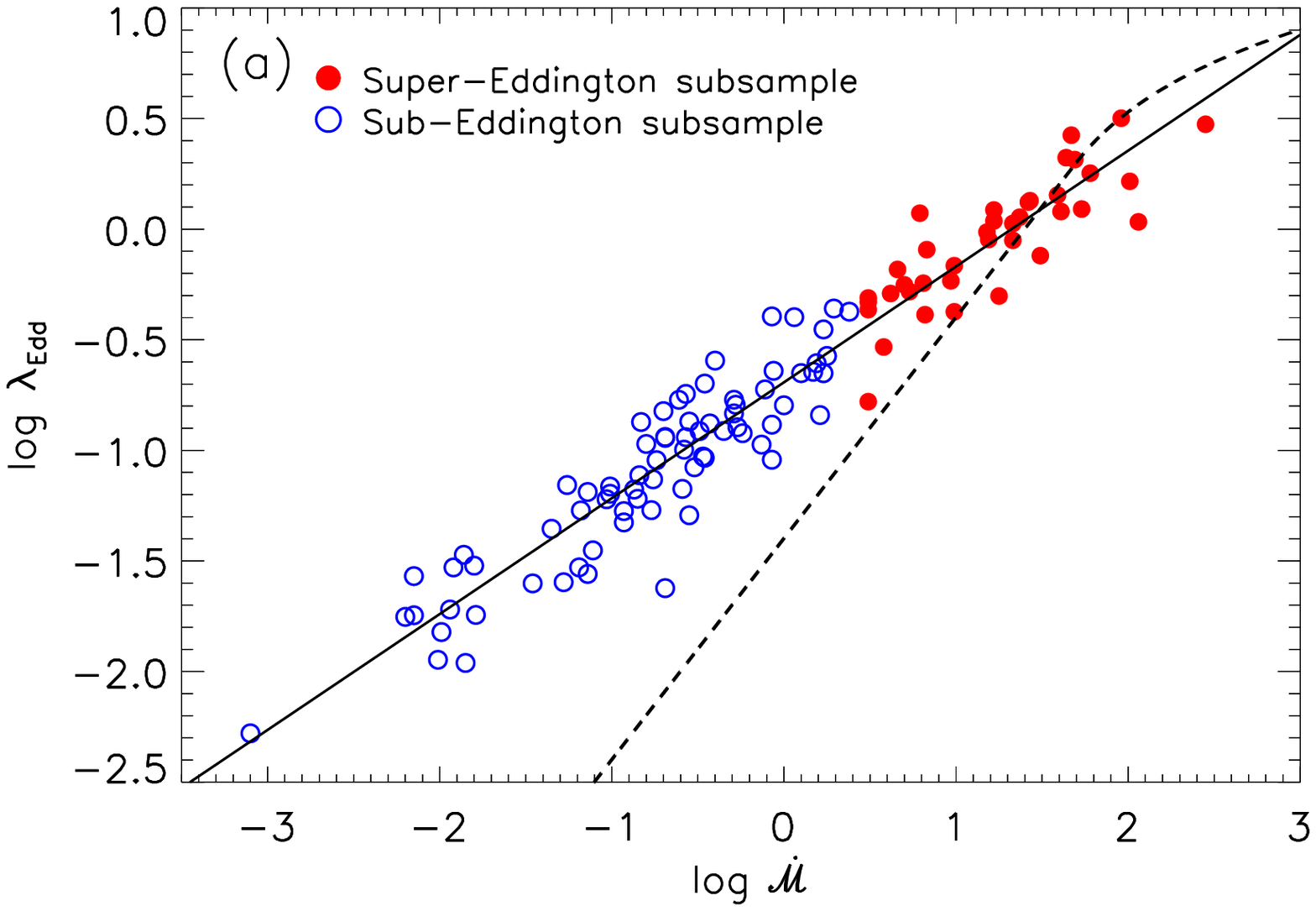}
\includegraphics[scale=0.50]{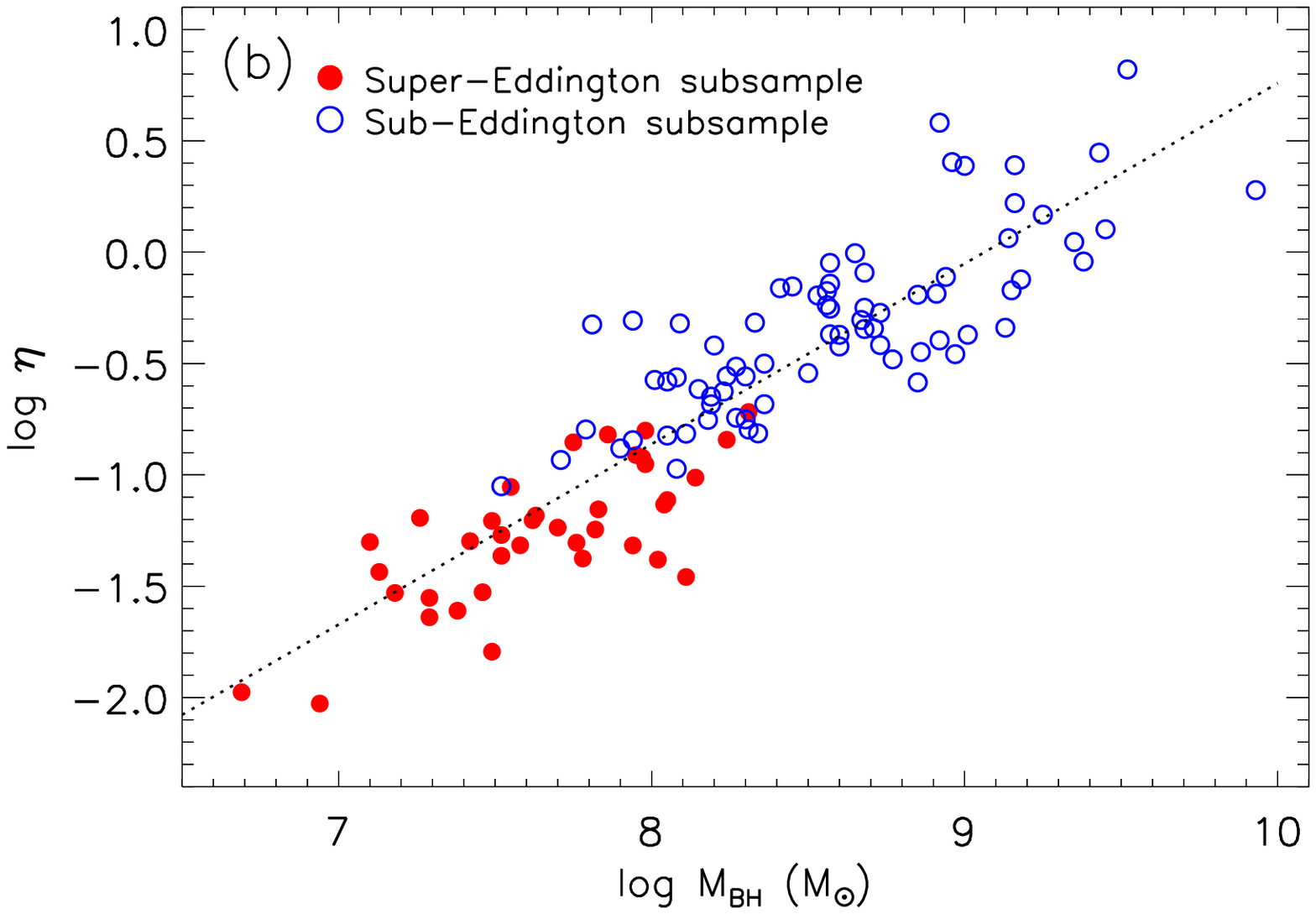}
}
\caption{{\bf (a):} \newedd\ vs. \mdot\ for the super- and \hbox{sub-Eddington} subsamples.
The solid line is the correlation between \edd\ and \mdot, with a \hbox{power-law} slope of $0.52$.
The dashed lines is the \mdot--\edd\ correlation in \cite{Mineshige2000} with $\eta=0.04$ in the sub-Eddington regime.
 {\bf (b):} $\eta$ vs. \mbh\ for the \hbox{super-} and \hbox{sub-Eddington} subsamples.
The dotted line is the correlation between $\eta$ and \mbh, with a \hbox{power-law} slope of $0.81$.
}

\label{fig-eddvsmdot}
\end{figure*}

\subsection{Estimation of \hbox{Black-Hole} Masses} \label{sec:bhmass}

We estimate the \hbox{black-hole} masses adopting the virial mass formula \hbox{$M_{\rm BH}=f{{V^{2}}_{\rm FWHM}}R_{\rm H\beta}/G$}, where $f$ is the virial factor, ${V_{\rm FWHM}}$ is the FWHM of the broad ${\rm H\beta}$ emission line, $G$ is the gravitational constant,
and $R_{\rm H\beta}$ is the \hb\ BLR size \citep[e.g.,][]{Kaspi2000,Kaspi2005,Bentz2013}.
We adopt $f=1$ following \cite{du2019}.

The updated \hbox{$R$--${L}$} relation from \citep{du2019} can be expressed as
\begin{equation}\label{1}
{\rm log}{R_{\rm H\beta}}={1.64_{-0.06}^{+0.06}}+{0.45_{-0.04}^{+0.04}}{\rm log}{l_{44}}+{-0.35_{-0.09}^{+0.09}}{\cal{R}_{\rm Fe}},
\end{equation}
where ${\cal{R}_{\rm Fe}}$ is the relative strength of the optical \ion{Fe}{2} line emission (see Footnote~10) and $l_{44}$ is the 5100~{\AA} luminosity in units of $10^{44}~{\rm erg}{~\rm s}^{-1}$.
This relation indicates that the BLR sizes for \hbox{super-Eddington} accreting AGNs are relatively smaller than that of \hbox{sub-Eddington} accreting AGNs, as \hbox{super-Eddington} accreting AGNs usually have larger ${\cal{R}_{\rm Fe}}$ values \citep{boroson1992, hu2008, dong2011}.
With the updated \hbox{$R$--${L}$} relation, the \hbox{single-epoch} virial mass formula can be expressed as
\begin{equation}\label{2}
  {\rm log}\left(\frac{M_{\rm BH}}{M_{\odot}}\right)=7.83+2{\rm log}\left({V}_{\rm H\beta}\right)+0.45{\rm log}\left(l_{46}\right)-0.35{\cal{R}_{\rm Fe}},
\end{equation}
where ${V}_{\rm H\beta}={\rm FWHM}_{\rm H\beta}/{\rm 10^{3}~km~s^{-1}}$, and $l_{46}$ is the 5100~{\AA} luminosity in units of $10^{46}~{\rm erg}{~\rm s}^{-1}$.
We list the \mbh\ values and other optical properties for the 113 quasars in our final sample in Table~\ref{tbl-optical}.
Our final sample has \mbh\ values ranging from $10^{6.7}$ to $10^{9.9}~{M_{\odot}}$, with a median value of $10^{8.2}{M_{\odot}}$.
The \mbh\ values are used to derive the \mdot\ and \edd\ values for our final sample (see Section~\ref{sec:ar} below).

For comparison, we also use the conventional \hbox{$R$--${L}$} relation calibrated by \cite{Kaspi2005} to estimate the \hbox{black-hole} masses.
\cite{Netzer2007} used the \hbox{$R$--${L}$} relation calibrated by \cite{Kaspi2005}, and obtained a virial mass formula expressed as
\begin{equation}\label{3}
  {\rm log}\left(\frac{M_{\rm BH,NT}}{M_{\odot}}\right)=8.02+2{\rm log}\left({V}_{\rm H\beta}\right)+0.65{\rm log}\left(l_{46}\right).
\end{equation}
We use this formula to obtain \tnmbh\ values, and then use the \tnmbh\ values to derive \tnedd\ and \mdotold\ values.
A comparison of the \hbox{black-hole} masses estimated using the two \hbox{$R$--${L}$} relations is shown in Figure~\ref{fig-mass}a.
The \tnmbh$/$\mbh\ ratios have a mean value of 1.87 for the \hbox{super-Eddington} subsample, and a mean value of 1.09 for the \hbox{sub-Eddington} subsample.
The comparison shows that the two sets of \hbox{black-hole} masses differ mainly for \hbox{super-Eddington} accreting quasars, where the \hbox{black-hole} masses estimated from Equation~\ref{2} are generally smaller.

\begin{deluxetable*}{lccccccccrrrr}
\tabletypesize{\scriptsize}
\tablewidth{0pt}
\tablecaption{Optical and Radio Properties for the Final Sample}
\tablehead{
\colhead{Object Name}   &
\colhead{Redshift}   &
\colhead{$M_{\rm i}$}   &
\colhead{log ${\rm FWHM}_{\rm H\beta}$}    &
\colhead{log ${L}_{\rm H\beta}$}    &
\colhead{log ${L}_{\rm Fe}$}   &  
\colhead{log ${L}_{\rm 5100}$}   &
\colhead{log ${L}_{\rm Bol}$}   &
\colhead{log \mbhtd}   &
\colhead{log \newedd}   &
\colhead{log \mdotnew}   & 
\colhead{$R$}   &   \\
\colhead{(J2000)}   &
\colhead{ }   &
\colhead{[$z$=2]}   &
\colhead{(${\rm km}~{\rm s}^{-1}$)}    &
\colhead{(${\rm erg}~{\rm s}^{-1}$)}    &
\colhead{(${\rm erg}~{\rm s}^{-1}$)}   &  
\colhead{(${\rm erg}~{\rm s}^{-1}$)}   &
\colhead{(${\rm erg}~{\rm s}^{-1}$)}   &
\colhead{($M_{\odot}$)}   &
\colhead{ }   &
\colhead{ }   &
\colhead{ }   &    \\
\colhead{(1)}          &
\colhead{(2)}          &
\colhead{(3)}          &
\colhead{(4)}          &
\colhead{(5)}          &
\colhead{(6)}          &
\colhead{(7)}          &
\colhead{(8)}          &
\colhead{(9)}          &
\colhead{(10)}        &
\colhead{(11)}         &
\colhead{(12)}
}
\startdata
002233.27-003448.4 & $0.504 $& $-23.78 $& $3.26 $& $42.81 $& 
$42.67 $& $44.51 $& $45.64 $& $7.42 $& $0.12 $& $1.42 $&
$<0.99 $\\
004319.74+005115.4 & $0.309 $& $-23.79 $& $3.97 $& $42.63 $& 
$41.96 $& $44.43 $& $45.63 $& $9.00 $& $-1.47 $& $-1.86 $&
$3.30 $\\
005709.94+144610.1 & $0.172 $& $-24.47 $& $4.00 $& $43.09 $& 
$-1.00 $& $44.94 $& $45.71 $& $9.35 $& $-1.74 $& $-1.79 $&
$2.18 $\\
012549.97+020332.2 & $0.500 $& $-23.77 $& $3.78 $& $42.93 $& 
$-1.00 $& $44.79 $& $45.73 $& $8.85 $& $-1.22 $& $-1.03 $&
$<2.46 $\\
013418.19+001536.7 & $0.401 $& $-25.34 $& $3.72 $& $43.37 $& 
$42.95 $& $45.18 $& $46.10 $& $8.77 $& $-0.77 $& $-0.29 $&
$<0.76 $\\
014959.27+125658.0 & $0.432 $& $-24.33 $& $3.54 $& $43.00 $& 
$42.63 $& $44.74 $& $45.84 $& $8.19 $& $-0.45 $& $0.23 $&
$<4.17 $\\
015950.24+002340.8 & $0.163 $& $-24.04 $& $3.45 $& $42.78 $& 
$42.77 $& $44.77 $& $45.77 $& $7.83 $& $-0.16 $& $0.99 $&
$6.10 $\\
020011.52-093126.2 & $0.360 $& $-23.81 $& $3.90 $& $43.35 $& 
$42.70 $& $45.06 $& $45.70 $& $9.13 $& $-1.53 $& $-1.19 $&
$<0.95 $\\
020039.15-084554.9 & $0.432 $& $-24.54 $& $3.25 $& $42.81 $& 
$42.78 $& $44.81 $& $45.81 $& $7.46 $& $0.25 $& $1.78 $&
$<1.04 $\\
020354.68-060844.0 & $0.464 $& $-24.25 $& $3.80 $& $43.18 $& 
$42.24 $& $44.95 $& $45.69 $& $8.92 $& $-1.33 $& $-0.93 $&
$<2.28 $\\
020840.66-062716.7 & $0.092 $& $-20.73 $& $3.62 $& $41.32 $& 
$41.14 $& $43.49 $& $44.19 $& $7.71 $& $-1.62 $& $-0.69 $&
$<2.04 $
\enddata
\tablecomments{Column (1): Name of the object, in order of increasing right ascension; Column (2): Redshift; Column (3): The absolute magnitude in the $i$-band at $z=2$; Column (4): Logarithm of the FWHM of the broad ${\rm H\beta}$ emission line in units of ${\rm km}~{\rm s}^{-1}$; Column (5): Logarithm of the luminosity of the ${\rm H\beta}$ broad emission line in units of ${\rm erg}~{\rm s}^{-1}$; Column (6): Logarithm of the luminosity of the optical \ion{Fe}{2} line emission in units of ${\rm erg}~{\rm s}^{-1}$, which is labeled as `$-1.00$' when there is no \ion{Fe}{2} component measured from the optical spectrum; Column (7): Logarithm of the continuum luminosity at \hbox{rest-frame} 5100 {\AA} in units of ${\rm erg}~{\rm s}^{-1}$; Column (8): Logarithm of the bolometric luminosity derived from integrating the SED in units of ${\rm erg}~{\rm s}^{-1}$; Column (9): Logarithm of the \hbox{black-hole} mass derived from Equation~\ref{2}, in units of solar mass; Column (10): Logarithm of the Eddington ratio; Column (11): Logarithm of the dimensionless accretion rate; Column (12): The radio loudness parameter or its upper limit.\\
(This table is available in its entirety including 113 objects in a machine-readable form in the online journal. A portion is shown here for guidance regarding its form and content.)}
\label{tbl-optical}
\end{deluxetable*}

\subsection{Dimensionless Accretion Rates and Eddington Ratios} \label{sec:ar}

We estimate the dimensionless accretion rates and Eddington ratios for our final sample.
For each quasar in our sample, the \mdot\ value is estimated based on the standard thin disk accretion model \citep[e.g.,][]{wang2014a, du2016} and can be expressed as
\begin{equation}\label{4-0} 
\dot{\mathscr{M}}=20.1({l_{44}}/{\cos i})^{3/2}{m_{7}}^{-2},
\end{equation}
where ${m_{7}}={M_{\rm BH}}/{10^{7}M_{\odot}}$.
We adopt ${\cos i}=0.75$ in this study (see \citealt{du2016} for discussions). 

We note that the above formula is likely also valid for estimating the dimensionless accretion rates for \hbox{super-Eddington} accreting AGNs where thick accretion disks are generally expected, and it has been frequently adopted in recent reverberation mapping studies of \hbox{super-Eddington} accreting AGNs \citep[e.g.,][]{du2014,du2016,hu2015,wangf2016,Li2018}.
Theoretically, the slim disk model \citep{wang1999b,wang1999} indicates that the 5100~\AA\ disk emission region is beyond the photon trapping radius provided that \mdot$~\lesssim3\times10^{3}{m_{7}}^{-1/2}$ (see Footnote 8 of \citealt{du2016}), and thus a standard thin disk solution still applies when adopting the 5100~\AA\ luminosity to estimate the dimensionless accretion rate.
Among the 113 quasars in our final sample, none has the dimensionless accretion rate exceeding the above limit.
Observationally, studies on the spectral energy distributions (SEDs) of \hbox{super-Eddington} accreting AGNs often found that their optical/UV SEDs are well fit by the standard thin disk model, and any thick disk emission signature likely only exists in the extreme UV (EUV) where few observational data are available \citep[e.g.,][]{Castell2016,Kubota2019}.
In addition, from a recent {\it Swift} \hbox{accretion-disk} reverberation mapping campaign on the \hbox{super-Eddington} accreting AGN \hbox{Mrk 142} (Cackett E.\ et al.\ submitted), multiwavelength time lags in the optical/UV were found to follow
in general the $\tau(\lambda)\propto\lambda^{4/3}$ relation that is consistent with the thin disk model, suggesting that the optical/UV emission is likely still from a thin disk.
Therefore, we use Equation~\ref{4-0} to estimate the dimensionless accretion rates for all our sample quasars, and the obtained \mdot\ values are listed in Table~\ref{tbl-optical}.
The dimensionless accretion rates for the final sample range from $7.9\times10^{-4}$ to 280, with a median value of 0.54.

We adopt a criterion of \mdot\ $> 3$ to select \hbox{super-Eddington} accreting quasars \citep[e.g.,][]{wang2014a,du2016}.
Based on this criterion, 38 (34\%) of the 113 quasars in our final sample are considered \hbox{super-Eddington} accreting quasars, and we refer to these quasars as `the \hbox{super-Eddington} subsample'.
The other 75 quasars constitute the \hbox{sub-Eddington} subsample.
We show the ${\cal{R}_{\rm Fe}}$ vs. \mbh\ distributions for the super- and \hbox{sub-Eddington} subsamples in Figure~\ref{fig-mass}b.
The \hbox{super-Eddington} subsample has larger ${\cal{R}_{\rm Fe}}$ values on average, which is consistent with previous findings \citep[e.g.,][]{boroson1992, hu2008, dong2011}.

We estimate the bolometric luminosities for our sample quasars by integrating their SEDs.
We collect their near infrared (NIR), optical, and UV photometric data from the public catalogs of the Two Micron All Sky Survey (2MASS; \citealt{Skrutskie2006}), SDSS, and {\it Galaxy Evolution Explorer\/} ({\it GALEX\/}; \citealt{Martin2005}).
We correct the SED data of each quasar for the Galactic extinction at its source position.
Among the 113 quasars in our final sample, 24 quasars do not have 2MASS photometric data, and 13 quasars do not have {\it GALEX\/} photometric data.
We construct the SEDs following mainly the procedure described in Section 3.1 of \cite{Davis2011}.
The SEDs between \hbox{1 \micron} and 1549~\AA\ are simple linear interpolations from the \hbox{NIR-to-UV} photometric data; for the 24 quasars without 2MASS data, the NIR SEDs are linear extrapolations from the SDSS data adopting a \hbox{power-law} spectral slope ($F_{\upsilon}\propto\upsilon^{\alpha}$) of $-0.3$ \citep{Davis2011}, and for the 13 quasars without {\it GALEX\/} data, the UV SEDs are linear extrapolations from the SDSS data adopting a spectral slope of $-0.5$ \citep{Vanden2001}.
The \hbox{1--30 \micron} SEDs are set to power laws with a spectral slope of $1/3$ \citep{Davis2011}.
We then add the \hbox{EUV-to-\xray} SEDs.
We assume a spectral slope of $-1$ between 1549~\AA\ and 1000~\AA.
The \hbox{2--10 keV} \hbox{power-law} spectra are obtained from our spectral fitting in Section~\ref{sec:dataanalysis} below, with spectral slopes of $1-$\gam.
We estimate the \hbox{0.2--2 keV} spectral slopes from the \hb\ FWHM using the relation in \cite{Brandt2000}.
The spectra between 1000~\AA\ and \hbox{0.2 keV} are then simple power laws connecting the two endpoints.
We integrate the \hbox{30 \micron--10 keV} SEDs to obtain the bolometric luminosities (\lbol), and the derived values are listed in Table~\ref{tbl-optical}.
For our sample, the \lbolsed\ values range from $1.5\times10^{44}~{\rm erg}~{\rm s}^{-1}$ to $5.2\times10^{46}~{\rm erg}~{\rm s}^{-1}$, with a median value of $4.1\times10^{45}~{\rm erg}~{\rm s}^{-1}$.

We caution that there may be considerable uncertainties associated with the bolometric luminosities derived from the multiwavelength SEDs above, especially for the  \hbox{super-Eddington} accreting quasars.
Besides potential \hbox{host-galaxy} contaminations in the \hbox{NIR--optical} SEDs that are usually small for luminous quasars and potential variability effects due to the \hbox{non-simultaneous} SED data, \hbox{super-Eddington} accreting quasars might have significantly enhanced EUV emission compared to typical quasars \citep[e.g.,][]{Davis2011,Jin2012,Castell2016,Kubota2019}.
There is no clear observational constraint on the EUV emission from \hbox{super-Eddington} accreting quasars due to the lack of data, and thus we cannot evaluate the scale of such a potential bias on the \lbol\ values for the \hbox{super-Eddington} subsample.

Another common approach of obtaining bolometric luminosities is through
the use of bolometric corrections to the optical luminosities (e.g., $L_{\rm Bol,BC}{\approx}{k_{\rm Bol}}{L}_{\rm 5100}$).
We also estimate the bolometric luminosities for our sample quasars using bolometric corrections. We first adopt the bolometric correction factors in \cite{Netzer2019}, which are expressed as
\begin{equation}\label{4} 
{k_{\rm Bol}}=40\left(L_{5100}/10^{42}~{\rm erg}{~\rm s}^{-1}\right)^{-0.2}.
\end{equation}
The derived \lbolbc\ values are comparable to our SED derived \lbolsed\ values, and the \lbolbc\ to \lbolsed\ ratios range from $0.53$ to $4.04$, with a median value of $1.17$.
It appears that the agreement between the two sets of estimates is slightly better for the \hbox{super-Eddington} subsample, with the median value of \lbolbc\ to \lbolsed\ ratios being $1.05$, while it is $1.23$ for the \hbox{sub-Eddington} subsample.
Next, we adopt a constant bolometric factor, that was often found from observations on large AGN samples \citep[e.g.,][]{,Richards2006,Duras2020}, to obtain another set of \lbolbc\ estimates.
We set ${k_{\rm Bol}}$ to be 10 following \cite{Kaspi2000}.
The derived \lbolbc\ values are again comparable to our SED derived \lbolsed\ values.
The \lbolbc\ to \lbolsed\ ratios range from $0.34$ to $2.28$, with a median value of $1.01$; the median values are 0.95 for the \hbox{super-Eddington} subsample and $1.08$ for the \hbox{sub-Eddington subsample}.
We caution that the above bolometric corrections were derived from thin disk models or from typical AGN SEDs, and they are probably still not applicable in the \hbox{super-Eddington} regime, where larger correction factors are likely expected \citep[e.g.,][]{Castell2016,Netzer2019}.
We consider that the differences between the two sets of \lbolbc\ values and the SED derived \lbol\ values simply reflect the systematic offsets between the different methods used to estimate the bolometric luminosities and they do not provide useful insight into the accuracy of the SED derived \lbolsed\ values in the \hbox{super-Eddington} regime.
In our following analysis, we adopt the SED derived \lbol\ values; using either set of the \lbolbc\ values above instead would not change the results significantly.

With the Eddington luminosities computed as \hbox{$L_{\rm Edd}={\rm 1.26}{\times}{\rm 10^{38}}({M_{\rm BH}}/{M_{\odot}})~{\rm erg~s^{-1}}$}, the Eddington ratios for our sample quasars are derived.
We list the Eddington ratios for our final sample in Table~\ref{tbl-optical}, which range from $5.2\times10^{-3}$ to $3.3$, with a median value of $0.16$.
Compared to the dimensionless accretion rates, the Eddington ratios have additional uncertainties associated with the bolometric luminosities that are especially uncertain for the \hbox{super-Eddington} subsample. Therefore, we focus our study on the \hbox{\gam--\mdot} correlation below. We still keep the analysis of the \hbox{\gam--\edd} correlation, mainly for the comparisons of this correlation to those found in previous studies.

\subsection{The Connection Between \mdot\ and \edd} \label{sec:edd-mdot}

Analytical solutions indicate that for a \hbox{black-hole} with given \hbox{black-hole} mass and spin, there is a connection between the dimensionless accretion rate and the Eddington ratio \citep[e.g.,][]{Mineshige2000,Watarai2000}.
When the black hole is \hbox{sub-Eddington} accreting, the Eddington ratio should change linearly with the dimensionless accretion rate because the radiative efficiency is a constant for a standard thin disk.
When the accretion is in the \hbox{super-Eddington} regime, the radiative efficiency decreases significantly due to the photon trapping effect, which indicates that the linear correlation between \edd\ and \mdot\ disappears.

In this study, we investigate the correlation between \edd\ and \mdot\ for our sample objects.
The distribution of the \edd\ versus \mdot\ values for our sample is shown in Figure~\ref{fig-eddvsmdot}a.
For the full sample, there is a significant \hbox{power-law} correlation between \edd\ and \mdot, with a \hbox{power-law} slope of 0.52.
The \hbox{\edd--\mdot} correlation from a semi-analytical model (\citealt{Mineshige2000}, \citealt{Watarai2000}; $\eta=0.04$ in the radiatively efficient case) is shown in Figure~\ref{fig-eddvsmdot}a for comparison, where \edd\ changes linearly with \mdot\ (with a \hbox{power-law} slope of 1) when \mdot$<50$, and it saturates above \mdot$=50$.
The strong \hbox{\edd--\mdot} correlation and its deviation from the analytical expectation can be understood from Equation \ref{4-0}, which indicates \hbox{\mdot~$\propto{L_{5100}}^{1.5}{M_{\rm BH}}^{-2}$}.
The \edd\ parameter is related to $L_{5100}$ and \mbh\ in the form: \hbox{\edd~$\propto{L_{\rm Bol}}{M_{\rm BH}}^{-1}\propto{L_{5100}}{M_{\rm BH}}^{-1}$}, considering that the bolometric luminosity is generally linearly scaled to the optical luminosity with some scatter.
Therefore, \edd\ is correlated to \mdot\ with a \hbox{power-law} form: \hbox{$\lambda_{\rm Edd}\propto{L_{5100}}^{0.25}{\dot{\mathscr{M}}}^{0.5}$}.
The dependence on ${L_{5100}}$ is small and the range of ${L_{5100}}$ for our sample quasars is also limited.
This explains the \hbox{$\lambda_{\rm Edd}\propto{\dot{\mathscr{M}}}^{0.52}$} \hbox{power-law} correlation we observe in Figure~\ref{fig-eddvsmdot}a.

This \edd--\mdot\ relation also indicates that the radiative efficiency $\eta$ is correlated with \hbox{black-hole} mass in the form of $\eta=\lambda_{\rm Edd}/{\dot{\mathscr{M}}}\propto{L_{5100}}^{-0.5}{{M}_{\rm BH}}$.
The $\eta$ paramter is roughly linearly correlated with \mbh.
In Figure~\ref{fig-eddvsmdot}b, we plot the best-fit \hbox{$\eta$--\mbh} correlation for our final sample.
The best-fit \hbox{$\eta$--\mbh} correlation is \hbox{$\eta\propto {{M}_{\rm BH}}^{0.81}$}.
Previous studies also found similar correlations between $\eta$ and \mbh\ \citep{Davis2011,Chelouche2013}.
For example, \citep{Davis2011} used a sample of PG quasars and found that \hbox{$\eta\propto {{M}_{\rm BH}}^{0.53}$}.
We compare the \mdot\ values derived from Equation \ref{4-0} to those derived from Equation 7 in \cite{Davis2011}, and they are consistent.
We explore several possible factors that may explain the unusual \edd--\mdot\ correlation and the deviation from the analytical expectation:
\begin{enumerate}
\item
The standard thin disk model needs to be modified, and thus the computation of \mdot\ using Equation 4 is not appropriate (e.g., see Section 4.3 of \citealt{Davis2011}).
\item
The bolometric luminosities for the \hbox{super-Eddington} accreting quasars are highly uncertain and may even be biased.
However, we note that if considering only the \hbox{sub-Eddington} subsample, the strong nonlinear \hbox{$\lambda_{\rm Edd}\propto{\dot{\mathscr{M}}}^{0.52}$} correlation still exists and deviates from the theoretical expectation (Figure~\ref{fig-eddvsmdot}a).
Therefore, we consider that the uncertainties on the bolometric luminosities are not the main cause for the unusual \hbox{\edd--\mdot} correlation.
\item
There is a real connection between the radiative efficiency and the \hbox{black-hole} masses (Figure~\ref{fig-eddvsmdot}b).
If $\eta$ increases as \mbh\ increases, the \hbox{\edd--\mdot} correlation ($\eta=$\edd$/$\mdot) would be flatter than the power law with a unity slope, as smaller \mdot\ values should correspond to larger \mbh\ and thus larger $\eta$ values.
Such a $\eta$--\mbh\ correlation would suggest that the \hbox{black-hole} spin increases as the black hole gains its mass via accretion \citep[e.g.,][]{Davis2011}. 
\end{enumerate}

In addition, the uncertainties on the \mbh\ and \lbolsed\ parameters may also contribute partially to the $\eta$--\mbh\ correlation in Figure~\ref{fig-eddvsmdot}b (see Sections 4.1 and 4.2 of \citealt{Davis2011}) and thus the \edd--\mdot\ correlation in Figure~\ref{fig-eddvsmdot}a.
Nevertheless, it is possible that some of the above points are working together to create the observed \edd--\mdot\ correlation, and it appears inevitable to obtain such a correlation if the standard thin disk model is adopted.
In the current study, we focus on the relation between the hard \xray\ photon index and the accretion rate for super-Eddington accreting quasars.
Considering the additional uncertainties associated with the bolometric luminosities, we prioritize the use of the \mdot\ to represent the accretion rates for our sample objects.

\section{Data Analysis}\label{sec:data}
\subsection{\chandra\ Data Analysis}\label{sec:cdata}
For each \chandra\ observation, we analyze the data using the \chandra\ Interactive Analysis of Observation\footnote{https://cxc.harvard.edu/ciao/.} (CIAO; v4.10) tools.
We first use the {\sc chandra\_repro} script to generate a new level 2 event file, and then filter background flares by running the {\sc deflare} script using an iterative 3${\sigma}$ clipping algorithm to obtain the cleaned event file.
We create an \xray\ image in the \hbox{0.5--7~keV} band from the cleaned event file by running the {\sc dmcopy} tool.

To search for \xray\ sources in the \xray\ image, we use the {\sc wavdetect} tool \citep{Freeman2002} with a false-positive probability threshold of $10^{-6}$ and scale sizes of 1, 1.414, 2, 2.828, 4, 5.656, 8 pixels.
We then match the optical position of the quasar to the \xray\ source positions to search for the \xray\ counterpart, using a 3\arcsec\ matching radius.
For the quasars in our final sample, the offsets between the optical positions and the \xray\ counterparts positions have a mean value of 0.99\arcsec.
To extract the source spectrum for each source, we choose a circular source region centered on the \xray\ counterpart, with a radius of the 90\% PSF size plus 3\arcsec
; we use the \hbox{{\sc psfsize\_src}} script to obtain the size of the PSF with 90$\%$ enclosed counts at \hbox{1.5~keV}.
To extract the background spectrum, we choose an annulus region with radii of three times and five times the source extraction radius.
The background region is also centered on the \xray\ position of the quasar.
For three sources that are in crowded areas or near the chip edges, we make the source or background regions smaller to avoid contamination from other detected \xray\ sources or bias from the chip edges.
We use the {\sc specextract} tool to extract the \xray\ source spectrum.

In order to measure the hard \xray\ photon index for each \chandra\ source in our sample, we fit the \xray\ spectra in the \hbox{observed-frame} \hbox{$2/(1+z)$--7~keV} band.
To exclude the Fe K complex, we exclude the spectrum in the \hbox{rest-frame} \hbox{5.5--7.5 keV} band \citep{Brightman2013}.
We obtain the number of net source counts in the \hbox{observed-frame} \hbox{$2/(1+z)$--7~keV} band (excluding the Fe K complex) by subtracting the estimated number of background counts in the source aperture from the number of source counts.
The number of background counts is scaled using the area scaling factor which is the ratio between the areas of the background and source extraction regions.
The number of the net source counts is then used to select \xray\ bright quasars (see Section~\ref{sec:chandra}).

\begin{deluxetable*}{lcrcccccr}
\tabletypesize{\scriptsize}
\tablewidth{0pt}
\tablecaption{X-ray Properties for the Final Sample}
\tablehead{
\colhead{Object Name}   &
\colhead{Observatory}    &
\colhead{Observation}   &
\colhead{Cleaned}    &
\colhead{Net}   &
\colhead{$N_{\rm H,Gal}$}   &
\colhead{\gam\ }   &  
\colhead{log ${F}_{\rm 2-10 keV}$}   &
\colhead{$W$/d.o.f.}   \\
\colhead{(J2000)}   &
\colhead{ }    &
\colhead{ID}   &
\colhead{Exposure Time (ks)}    &
\colhead{Source Counts}   &
\colhead{($10^{20}~{\rm cm}^{-2}$)}   &  
\colhead{ }   &
\colhead{(${\rm erg}~{\rm cm}^{-2}~{\rm s}^{-1}$)}   &
\colhead{ }   \\
\colhead{(1)}          &
\colhead{(2)}          &
\colhead{(3)}          &
\colhead{(4)}          &
\colhead{(5)}          &
\colhead{(6)}          &
\colhead{(7)}          &
\colhead{(8)}          &
\colhead{(9)}                    
}
\startdata
002233.27-003448.4 &C &16226 &$64.7 $&$279.6 $&$2.78 $&
$2.22_{-0.16}^{+0.17}$& $-13.23 $&99.5/120 \\
004319.74+005115.4 &X &0090070201 &$16.2 $&$1839.9 $&$2.31 $&
$1.74_{-0.05}^{+0.05}$& $-12.02 $&613.7/735 \\
005709.94+144610.1 &C &865 &$4.7 $&$1079.8 $&$4.37 $&
$1.92_{-0.10}^{+0.10}$& $-11.38 $&162.3/204 \\
012549.97+020332.2 &X &0741300601 &$54.6 $&$427.0 $&$3.04 $&
$1.68_{-0.13}^{+0.13}$& $-12.28 $&302.9/326 \\
013418.19+001536.7 &C &7748 &$9.9 $&$292.7 $&$2.91 $&
$1.62_{-0.15}^{+0.16}$& $-12.28 $&99.5/138 \\
014959.27+125658.0 &X &0673770301 &$24.9 $&$315.6 $&$5.23 $&
$2.28_{-0.33}^{+0.43}$& $-12.85 $&166.8/327 \\
015950.24+002340.8 &C &4104 &$9.7 $&$856.3 $&$2.59 $&
$2.23_{-0.11}^{+0.12}$& $-11.88 $&162.7/193 \\
020011.52-093126.2 &C &15577 &$19.3 $&$465.9 $&$2.02 $&
$1.82_{-0.13}^{+0.13}$& $-12.41 $&123.1/159 \\
020039.15-084554.9 &C &6106 &$35.3 $&$231.7 $&$2.06 $&
$2.29_{-0.21}^{+0.21}$& $-12.96 $&102.1/126 \\
020354.68-060844.0 &X &0747190631 &$10.3 $&$218.0 $&$2.13 $&
$2.09_{-0.19}^{+0.19}$& $-12.68 $&182.8/226 \\
020840.66-062716.7 &X &0747190835 &$11.4 $&$253.0 $&$2.21 $&
$2.02_{-0.21}^{+0.21}$& $-12.46 $&206.5/247
\enddata
\tablecomments{Column (1): Name of the object, in order of increasing right ascension; Column (2): The observatory of the \xray\ observation; `C' represents \chandra\ and `X' represents \xmm; Column (3): The observation identifiers; Column (4): The cleaned exposure time; Column (5): The net counts in the \hbox{observed-frame} $2/(1+z)$--7~keV band for \chandra\ or $2/(1+z)$--10~keV band for \xmm, excluding the Fe K complex; Column (6): The Galactic extinction; Column (7): The hard \xray\ photon index; Column (8): The flux at the \hbox{observed-frame} 2--10 keV band in units of ${\rm erg}~{\rm cm}^{-2}~{\rm s}^{-1}$; Column (9): The W statistic value over the degree of freedom. \\
(This table is available in its entirety including 113 objects in a machine-readable form in the online journal. A portion is shown here for guidance regarding its form and content.)}
\label{tbl-xray}
\end{deluxetable*}

\subsection{\xmm\ Data Analysis}\label{sec:xdata}
We use the Science Analysis System (SAS; v1.2) for the \xmm\ data reduction.
We follow the standard procedure in the SAS Data Analysis Threads\footnote{https://www.cosmos.esa.int/web/xmm-newton/sas-threads.} to process the data.
For all sources, we only use the data from the PN camera.
We use the {\sc epproc} tool to get calibrated and concatenated event lists.
We use a count rate threshold of \hbox{0.4 cts/s} to filter background flares, and we use the {\sc tabgtigen} script to create good-time-intervals files.
We then use the {\sc evselect} tool to obtain the cleaned event files.

Based on the flare-filtered event files, we use the {\sc evselect} tool to construct images in the \hbox{$0.3$--$10$~keV} band.
Then we use the {\sc edetect\_chain} tool to detect point sources in the images.
For each quasar, we select a circular source region with a radius of 30\arcsec\ and a circular background region with a radius of 40\arcsec.
The source region is centered on the optical position of each quasar.
The background region is chosen to be on the same CCD chip as the source region, and is free of other \xray\ sources.
For six sources that are in crowded areas or near the chip edges, we make the source region smaller to avoid contamination from other detected \xray\ sources or bias from the chip edges.
We then use the {\sc evselect} tool to extract the \xray\ spectra in the \hbox{observed-frame} \hbox{0.1--10~keV} band.

To measure the hard \xray\ photon index \gam\ for each \xmm\ source in our sample, we fit the spectrum in the \hbox{observed-frame} \hbox{$2/(1+z)$--10~keV} band.
We also ignore the spectrum in the \hbox{rest-frame} \hbox{5.5--7.5 keV} band to exclude the Fe K complex.
We obtain the number of net source counts in the \hbox{observed-frame} \hbox{$2/(1+z)$--10~keV} band (excluding the Fe K complex) by subtracting from the number of source counts the estimated number of background counts in the source aperture.
Then we use the numbers of net source counts to select \xray\ bright quasars (see Section~\ref{sec:xmm}). 

\begin{figure}
\centerline{
\includegraphics[scale=0.50]{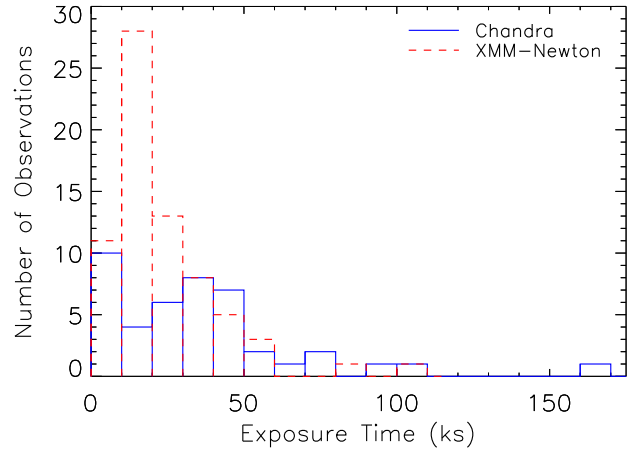}
}
\caption{Distributions of the cleaned exposure times of the \xray\ observations for our final sample, including 43 \chandra\ observations (blue solid line) and 70 \xmm\ observations (red dashed line).
~\\
}
\label{fig-exp}
\end{figure}

Among the 113 quasars in our final sample, there are six quasars which have both \chandra\ and \xmm\ observations.
We choose the observation which yields a larger number of net source counts (see~Section~\ref{sec:sdss}).
We adopt \chandra\ observations for 43 quasars (\chandra\ group) and \xmm\ observations for the other 70 quasars (\xmm\ group).
We show the histograms of the cleaned exposure times of the \xray\ observations for our final sample in Figure~\ref{fig-exp}.
The \xray\ properties for each quasar in our final sample, such as the cleaned exposure times and net source counts, are listed in Table~\ref{tbl-xray}.
For our final sample, the cleaned exposure times has a median value of \hbox{21.8 ks} and the numbers of net source counts have a median value of 465.9.

\begin{figure*}
\centerline{
\includegraphics[scale=0.50]{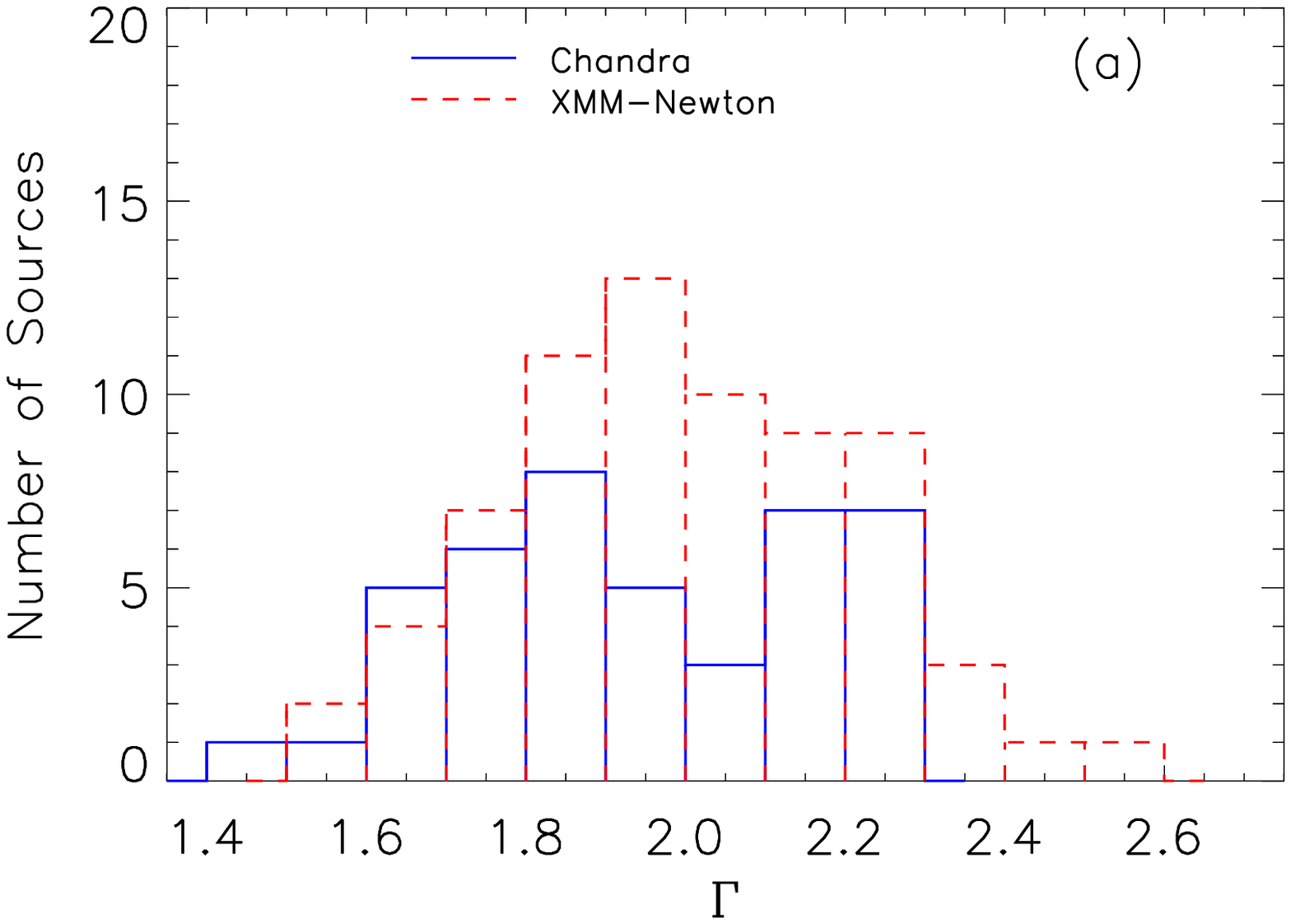}
\includegraphics[scale=0.50]{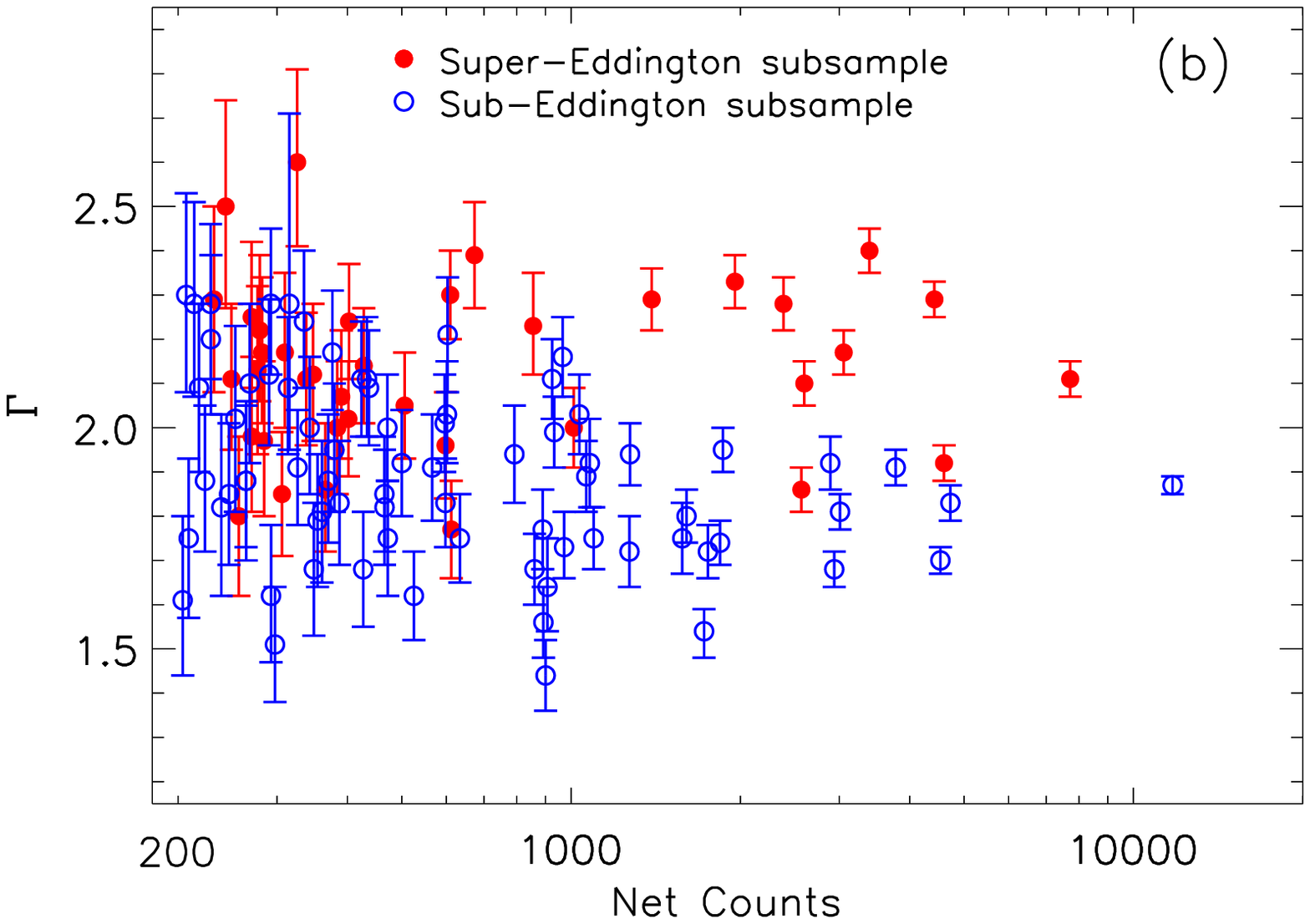}
}

\caption{{\bf (a):} Distributions of the \gam\ values for quasars in our final sample.
The blue solid line represents the distribution for the \chandra\ group and the red dashed line for the \xmm\ group.
{\bf (b):} The hard \xray\ photon index vs. net spectral counts for the super- and \hbox{sub-Eddington} subsamples.
}
\label{fig-counts}
\end{figure*}

\subsection{X-ray Spectral Fitting}\label{sec:dataanalysis}
To obtain the \gam\ value for each source, we fit the \xray\ spectrum.
We analyze the spectra of the 124 \hbox{radio-quiet} quasars (see Section~\ref{sec:rl}) using SPEX (v.3.05 \citealt{Kaastra1996}).
Following the guide of the SPEX cookbook\footnote{https://www.sron.nl/astrophysics-spex/manual.}, we use the {\sc trafo} tool in SPEX to convert the OGIP spectra into the SPEX format.
We group each \xray\ spectrum into at least one count per bin for spectral fitting, and we use the W statistic for parameter estimation. 

We use a redshifted ({\sc reds}) single \hbox{power-law} model ({\sc pow}) to fit the spectrum for each quasar. 
We consider the galactic absorption for each source by adding a neutral hydrogen gas absorption component ({\sc absm}).

As mentioned in the Section~\ref{sec:bal}, we aim to exclude \xray\ absorbed quasars in our sample via \xray\ spectral fitting.
Thus we add another redshifted absorption component to fit for any intrinsic absorption.
We use the {\sc ftest} tool in XSPEC (v.12.10.1; \citealt{Arnaud1996}) to evaluate whether the intrinsic absorption component is appropriate.
We identify five \xmm\ sources and six \chandra\ sources that have intrinsic absorption at a 95$\%$ confidence level, with ${N_{\rm H}}$ values in the range of ${\rm 6.4\times10^{21}}-{\rm 6.6\times10^{22}}$${\rm cm^{-2}}$.
We thus exclude these 11 quasars from our sample (see~Section~\ref{sec:bal}).
The spectra for the 113 quasars in our final sample are all fitted with a \hbox{power-law} model modified by Galactic absorption.

The C statistic \citep{Cash1979,Kaastra2017} in SPEX can provide us the confidence level of the spectral fitting results, while the W statistic is not able to do so.
Thus we also use the C statistic in the spectral fitting, and group the data using the optimal data bin size \citep{Kaastra2016}, which can be achieved via the {\sc obin} command in SPEX.
We compare the fitting results from the two different statistics.
We find that for the \chandra\ spectra, the \gam\ values measured from the W statistic + `one count per bin' is consistent with that from the C statistic + `obin'.
But for the \xmm\ spectra, especially those \xmm\ spectra with relatively smaller numbers of net source counts, the \gam\ values are not consistent.
The \xmm\ observations generally have more background counts than \chandra\ observations.
When the numbers of net source counts are relatively small compared to the number of background counts, the C statistic in SPEX may give bias results (see the SPEX Reference Manual\footnote{https://var.sron.nl/SPEX-doc/manualv3.05/manual.html.} for details).
Thus it is not suitable yet to use the C statistic in SPEX to fit the \xmm\ spectra with small numbers of net counts.
We still use W statistic + `one count per bin' results in this study.
The $W/$d.o.f values for our final sample range from 0.51 to 1.08, and have a median value of 0.86.

The distributions of the \gam\ values for our final sample is shown in Figure~\ref{fig-counts}a.
The \gam\ values of the \chandra\ group have a median value of 1.92, while the \gam\ values of the \xmm\ group have a median value of 1.99.
We perform the Kolmogorov-Smirnov test using the {\sc kstwo} tool in IDL.
The result shows that $D=0.21$ and $P=0.175$, which indicates that the \gam\ values of these two groups of sources are not statistically different.

We list the \xray\ properties from the spectral fitting for our final sample in Table~\ref{tbl-xray}.
We show the \gam\ values versus the net counts for the super- and \hbox{sub-Eddington} subsamples in Figure~\ref{fig-counts}b.
The uncertainties of the \gam\ values for our sample are generally smaller than 0.2, because we only select quasars with numbers of \xray\ net source counts larger than 200 (see Section~\ref{sec:chandra} and Section~\ref{sec:xmm}).

\section{Results}\label{sec:results}

The aim of our study is to investigate if there is any difference between the disk--corona connections in super- and \hbox{sub-Eddington} accreting quasars, by comparing the correlations between \gam\ and \mdot\ for these two types of accretion systems.
In this section, we examine the correlations between \gam\ and \mdot, and the correlations between \gam\ and \edd\ for the super- and \hbox{sub-Eddington} subsamples, respectively.
When performing the linear regression analysis, we consider the $1\sigma$  uncertainties of the \gam\ and log\mdot\ (log\edd) values.
We adopt the typical uncertainty of \mdot\ (\edd) to be \hbox{0.4 dex} (\hbox{0.2 dex}) from \cite{du2019}, that is dominated by the systematic uncertainty of deriving the \hbox{black-hole} mass from the updated \hbox{$R$--${L}$} relation and the virial mass formula.
We also investigate whether there is a correlation between \gam\ and \mbh.

\subsection{The Correlation Between \gam\ and \mdot} \label{sec:gamvsmdot}

For our \hbox{super-Eddington} subsample, we perform the Spearman rank correlation test using the {\sc r\_correlate} tool in IDL to investigate if there is a correlation between \gam\ and \mdot.
The result of the test is presented in Table~\ref{tbl-srlr}, which shows that the correlation between \gam\ and \mdot\ is statistically significant, with the null hypothesis probability $p=7.75\times{10}^{-3}$ and the Spearman rank coefficient $R_{\rm S}=0.43$. 

\begin{figure}
\centerline{
\includegraphics[scale=0.50]{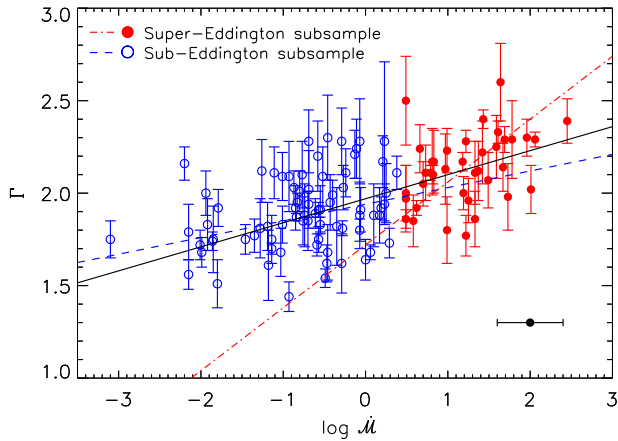}
}
\caption{The \hbox{\gam--\mdot} correlation for the 113 quasars in our final sample.
The blue dashed line and red dashed dotted line represent the correlations for our \hbox{sub-Eddington} and \hbox{super-Eddington} subsamples, respectively.
The black solid line represents the correlation for the 113 quasars in our final sample.
We show the size of the adopted uncertainties (0.4) of the log\mdot\ values in the lower right corner.
}

\label{fig-gamvsmdot}
\end{figure}

Then we use the {\sc linmix\_err} tool \citep{kelly2007} in the IDL Astronomy User's Library to perform the linear regression analysis.
The best-fit relation is 
\begin{equation}\label{5}
{\rm \Gamma}=(0.34\pm0.11){\rm log}{\dot{\mathscr{M}}}+(1.71\pm0.17).
\end{equation}
We list the parameters of the best-fit correlation in Table~\ref{tbl-srlr}, and we plot the \hbox{\gam--\mdot} correlation for the \hbox{super-Eddington} subsample in Figure~\ref{fig-gamvsmdot}.

We also perform the Spearman rank correlation test on the \hbox{sub-Eddington} subsample.
The \hbox{\gam--\mdot} correlation is statistically significant with the null hypothesis probability $p=9.98\times{10}^{-3}$.
However, the Spearman coefficient $R_{\rm S}$ value ($0.30$) is smaller than that of the \hbox{super-Eddington} subsample ($0.43$), suggesting that the \hbox{\gam--\mdot} correlation for the \hbox{sub-Eddington} subsample is weaker than that for the \hbox{super-Eddington} subsample.
We perform the linear regression analysis for the \hbox{sub-Eddington} subsample, the best-fit relation is 
\begin{equation}\label{6}
{\rm \Gamma}=(0.09\pm0.04){\rm log}{\dot{\mathscr{M}}}+(1.93\pm0.04).
\end{equation}
The slope of this correlation differs from that for the \hbox{super-Eddington} subsample (Equation~\ref{5}) at the $\approx2.1\sigma$ level.

We also investigate the \hbox{\gam--\mdot} correlation for all the 113 quasars in our final sample.
Using the same approaches above, we find that the \hbox{\gam--\mdot} correlation for the full sample is statistically significant, with $p=1.10\times{10}^{-10}$ and $R_{\rm S}=0.56$.
The best-fit relation is 
\begin{equation}\label{7}
{\rm \Gamma}=(0.13\pm0.02){\rm log}{\dot{\mathscr{M}}}+(1.97\pm0.02).
\end{equation}
We show the best-fit correlations between \gam\ and \mdot\ for the \hbox{sub-Eddington} subsample and full sample in Figure~\ref{fig-gamvsmdot}, and we list the best-fit parameters in Table~\ref{tbl-srlr}.

As mentioned in Section~\ref{sec:ar}, we also use the \tnmbh\ values to derive the \mdotold\ values.
We also investigate the correlation between \gam\ and \mdotold\ for our two subsamples and the full sample.
The \gam--\mdotold\ correlation (with $p=4.75\times{10}^{-2}$ and $R_{\rm S}=0.32$) for the \hbox{super-Eddington} subsample is also stronger than that for the \hbox{sub-Eddington} subsample (with $p=4.78\times{10}^{-2}$ and $R_{\rm S}=0.23$).
Generally, the \hbox{\gam--\mdot} correlations are slightly stronger than the \gam--\mdotold\ correlations, which may suggest that the \hbox{black-hole} masses estimated by the updated \hbox{$R$--${L}$} relation are more reliable.

\subsection{The Correlation Between \gam\ and \edd\ } \label{sec:gamvsedd}

We also perform the Spearman rank correlation tests on the two subsamples and the full sample to investigate the correlations between \gam\ and \edd.
The results of the tests are presented in Table~\ref{tbl-srlr}.
We find a significant correlation between \gam\ and \edd\ for the \hbox{super-Eddington} subsample (with $p=2.76\times10^{-4}$ and \hbox{$R_{\rm S}=0.56$}).
A weak correlation is found for the \hbox{sub-Eddington} subsample (with $p=7.10\times10^{-2}$ and \hbox{$R_{\rm S}=0.21$}).
We find a strong and statistically significant correlation between \gam\ and \edd\ for the full sample, with $p=1.32\times10^{-9}$ and $R_{\rm S}=0.53$.

\begin{figure}
\centerline{
\includegraphics[scale=0.50]{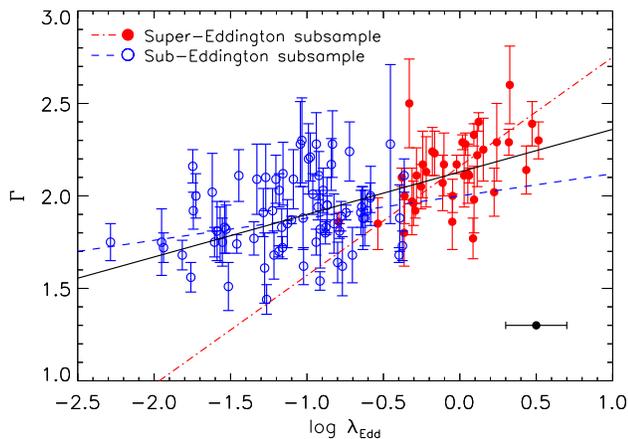}
}

\caption{The \gam--\newedd\ correlation for the 113 quasars in our final sample.
The blue dashed line and red dashed dotted line represent the correlations for our \hbox{sub-Eddington} and \hbox{super-Eddington} subsamples, respectively.
The black solid line represents the correlation for the 113 quasars in our final sample.
We show the size of the adopted uncertainties (0.2) of the log\edd\ values in the lower right corner.
}

\label{fig-gamvsedd}
\end{figure}

We use the {\sc linmix\_err} tool to perform the linear regression analysis.
The results are shown in Table~\ref{tbl-srlr}.
The best-fit correlation for the \hbox{super-Eddington} subsample is
\begin{equation}\label{8}
{\rm \Gamma}=(0.59\pm0.16){\rm log}{\lambda_{\rm Edd}}+(2.16\pm0.03),
\end{equation} 
while the best-fit correlations for the \hbox{sub-Eddington} subsample and the full sample are
\begin{equation}\label{9}
{\rm \Gamma}=(0.12\pm0.07){\rm log}{\lambda_{\rm Edd}}+(2.00\pm0.08),
\end{equation}
and  
\begin{equation}\label{10}
{\rm \Gamma}=(0.23\pm0.03){\rm log}{\lambda_{\rm Edd}}+(2.13\pm0.03).
\end{equation} 
We plot the \hbox{\gam--\edd} correlations for the two subsamples and the full sample in Figure~\ref{fig-gamvsedd}.
The results also show that the \hbox{\gam--log\edd} correlation slope for the \hbox{super-Eddington} subsample is steeper than that for the \hbox{sub-Eddington} subsample, which is consistent with the \hbox{\gam--log\mdot} correlation.
Due to the significant \hbox{power-law} correlation between \mdot\ and \edd\ (see Section~\ref{sec:edd-mdot}), the slopes of the \hbox{\gam--log\mdot} and the \hbox{\gam--log\edd} correlations are actually tightly connected.
For example, the slope of the \hbox{\gam--log\edd} correlation for the full sample are about twice the slope of the corresponding \hbox{\gam--log\mdot} correlation.

We compare our \hbox{\gam--\edd} correlation for the full sample to those in previous studies \citep[e.g.,][]{wang2004, Shemmer2008, Brightman2013}.
\cite{wang2004} found a correlation slope of $(0.26\pm0.05)$, \cite{Shemmer2008} found a correlation slope of $(0.31\pm0.01)$, and \cite{Brightman2013} found a correlation slope of $(0.32\pm0.05)$.
The \hbox{\gam--log\edd} correlation slope is $(0.23\pm0.03)$ for our full sample.
Considering the uncertainties, our correlation is generally consistent with those in previous studies; small differences might be caused by the different samples and methodologies utilized in these studies.

We also investigate the correlations between \gam\ and \tnedd.
The results are listed in Table~\ref{tbl-srlr}.
We still find that the \hbox{\gam--\edd} correlations are slightly stronger than the \hbox{\gam--\tnedd} correlations, and the
\hbox{\gam--log\tnedd} correlation slope for our \hbox{super-Eddington} subsample is steeper than that for our \hbox{sub-Eddington} subsample.
We notice that if we use \tnedd\ as the Eddington ratio for the linear regression, we can obtain a more consistent correlation with the previous studies, with the correlation slope of $(0.26\pm0.04)$ for the full sample.
The slope gets steeper because when we use the conventional \hbox{$R$--${L}$} relation, the \hbox{black-hole} masses of \hbox{super-Eddington} accreting quasars are larger, which lead to smaller Eddington ratios and thus a steeper slope of the correlation.

\subsection{The Correlation between \gam\ and Other Parameters} \label{sec:gamvsmbh}
The correlation between \gam\ and \mdot\ might be instead driven by potential correlations between \gam\ and other parameters.
The two parameters, ${L_{\rm 5100}}$ and \mbh, are used for calculating the dimensionless accretion rates and Eddington ratios.
We thus investigate whether there are any correlations between \gam\ and these two parameters.
 
For the full sample, we perform the Spearman rank correlation test on the \gam\ and ${L_{\rm 5100}}$ values, and we find no correlation between these two parameters, with $R_{\rm S} = 3.63{\times}{\rm 10^{-3}}$ and $p = 0.97$.
This also indicates that the \gam\ parameter has no apparent correlation with ${L_{\rm Bol}}$ which was derived from ${L_{\rm 5100}}$ (see Section~\ref{sec:ar}).
This result is consistent with the results of previous studies \citep[e.g.,][]{Shemmer2008,Risaliti2009,Brightman2013}.
We also find no correlations between \gam\ and ${L_{\rm 5100}}$ for the super- and \hbox{sub-Eddington} subsamples.

We also investigate the correlation between \gam\ and \mbh, and we show our results in Table~\ref{tbl-srlr}.
We find statistically significant, negative correlations between \gam\ and \mbh\ (and \tnmbh) for the full sample.
The \gam--\mbhtd\ correlation is stronger than the \gam--\tnmbh\ correlation.
For our \hbox{sub-Eddington} subsample, we also find a significant correlation between \gam\ and \mbh.
However, we find no correlation between \gam\ and \mbh\ for our \hbox{super-Eddington} subsample.

Previous studies also found correlations between \gam\ and \mbh.
For example, \cite{Risaliti2009} found that for their SDSS quasar sample, there is a negative correlation between \gam\ and \mbh.
They argued that due to the connection between \edd\ and \mbh, the partial degeneracy between the \hbox{\gam--\edd} correlation and \hbox{\gam--\mbh} correlation cannot be removed.
Similarly, for our full sample here, it is difficult to determine whether the \hbox{\gam--\mdot} or the \hbox{\gam--\mbh} correlation is the more fundamental correlation.
Nevertheless, we find a significant correlation between \gam\ and \mdot\ and no correlation between \gam\ and \mbh\ for our super-Eddington subsample, suggesting that the correlation between \gam\ and \mdot\ is more fundamental in the \hbox{super-Eddington} regime.

\section{Discussion}\label{sec:dis}

\subsection{The \hbox{\gam--\edd} Correlation}\label{sec:gamedddis}
Previous studies did not separate super- and \hbox{sub-Eddington} accreting AGNs in their samples, and thus the observed \hbox{\gam--\edd} correlation is probably a mixture of the different correlations of the two types of accretion systems.
In this study, we do find a strong and statistically significant \hbox{\gam--\edd} correlation for our full sample, and the slope of our correlation is consistent with those of previous studies (see Section~\ref{sec:gamvsedd}).

The difference between the slopes of the \hbox{\gam--log\edd} correlations for the two subsamples is large, with a slope of $(0.59\pm0.16)$ for the \hbox{super-Eddington} subsample and a slope of $(0.12\pm0.07)$ for the \hbox{sub-Eddington} subsample.
The slope is steeper for the \hbox{super-Eddington} subsample, suggesting that cooling of the corona (steepening of the \xray\ spectrum) is more efficient as \edd\ increases in the \hbox{super-Eddington} regime.
One natural explanation of such a phenomenon is that \edd\ is not a good representative of the accretion rate due to the photon trapping effect; for a give amount of change in \edd, the accretion rate actually changes by a larger amount.
In this scenario, the correlation between \gam\ and the accretion rate would be flatter for \hbox{super-Eddington} accreting quasars.
Such flattening is indeed observed in the \hbox{\gam--\mdot} correlation.
However, due to the complications of the \hbox{\edd--\mdot} correlation discussed in Section~\ref{sec:edd-mdot} and the uncertainties for deriving \lbol, we focus our discussion on the \hbox{\gam--\mdot} correlations below.

\subsection{The \hbox{\gam--\mdot} Correlation}\label{sec:gammdotdis}
We investigate the differences of the \hbox{\gam--\mdot} correlations in super- and \hbox{sub-Eddington} accreting quasars.
For our \hbox{super-Eddington} subsample, we find a statistically significant positive correlation between \gam\ and \mdot.
For our \hbox{sub-Eddington} subsample, we find a weaker positive \hbox{\gam--\mdot} correlation with a smaller $R_{\rm S}$ value and a larger p value, though the sample size of the \hbox{sub-Eddington} subsample is nearly twice that of the \hbox{super-Eddington} subsample and its dynamical range in \edd\ is also larger.
The correlation slope $(0.34\pm0.11)$ for the \hbox{super-Eddington} subsample is steeper than that $(0.09\pm0.04)$ for the \hbox{sub-Eddington} subsample, providing suggestive evidence that the \hbox{disk--corona} connections are different in these two types of accretion systems.

The steeper \hbox{\gam--\mdot} correlation in the \hbox{super-Eddington} regime might be an artificial effect caused by the soft \xray\ excess components in quasars spectra, which are usually quite strong in \hbox{super-Eddington} accreting quasars \citep[e.g.,][]{Boller1996,Kubota2019,Gliozzi2020}.
We fit the \xray\ spectra in the \hbox{rest-frame} \hbox{$>~2$ keV} band to reduce the contamination from possible soft \xray\ excess components.
But they may still contribute to the \hbox{rest-frame} \hbox{$>~2$ keV} spectra, leading to overestimated \gam\ values and a steeper \hbox{\gam--\mdot} correlation in the \hbox{super-Eddington} regime.
We examine such a possibility by fitting the spectra in the \hbox{rest-frame} \hbox{$>~3$ keV} band.
In this case, the number of our sample objects satisfying the criterion of $>200$ net counts reduces to 59, and only 19 of these are \hbox{super-Eddington} accreting quasars.
The updated \gam\ values for these 19 quasars differ slightly, with a median offset of $-0.06$.
After replacing these 19 \gam\ values, we perform the Spearman rank correlation test on the \hbox{super-Eddington} subsample to check whether the \hbox{\gam--\mdot} correlation still exists.
We find a weaker correlation with $R_{\rm S} = 0.36$ and $p = 0.025$, but this correlation is still stronger than that for the sub-Eddington subsample.
The new best-fit \hbox{\gam--log\mdot} correlation for the \hbox{super-Eddington} subsample has a slope of 0.34, consistent with the previous slope.
Therefore, we consider that the soft \xray\ excess does not contribute significantly to the steeper \hbox{\gam--\mdot} correlation for the \hbox{super-Eddington} subsample.

We note that a few recent studies have similar findings suggesting that the \gam\ versus accretion rate correlation is steeper in the super-Eddington regime (\citealt{Gliozzi2020}; Liu, H et al. in prep.).
If such a trend is indeed physical, it would suggest that cooling of the corona (steepening of the \xray\ spectrum) is more efficient as accretion rate increases in \hbox{super-Eddington} accreting AGNs.
The cooling of the corona is dominated by optical/UV seed photons from the accretion disk, and an increase of the photon flux received by the corona could enhance its cooling.
Considering that one main difference between a \hbox{super-Eddington} accreting disk and a \hbox{sub-Eddington} one is the thickness of the disk, one possible scenario is that disk photons are more easily to escape from the inner part of a thick disk because of the longer diffusion timescale in the thick disk as well as the stronger vertical advection in the inner region from effects such as magnetic buoyancy \citep[e.g.,][]{Jiang2014}.
The corona that is considered to be located in the immediate vicinity of the black hole \citep[e.g.,][]{Dai2010,Morgan2012,Luo2015,Kubota2018} thus receives a larger photon flux from a thick disk.
This qualitatively explains the steeper \hbox{\gam--\mdot} correlation in the \hbox{super-Eddington} regime.

Nevertheless, the \hbox{\gam--\mdot} correlations found in our study show large scatter.
A larger statistical sample is required to better constrain the correlations and confirm the difference between the \hbox{disk--corona} connections in super- and \hbox{sub-Eddington} accreting AGNS.
Such observational constraints will help us understand better the \hbox{super-Eddington} accreting systems that are still largely uncertain.

\section{Summary and Future Work}\label{sec:con}

\subsection{Summary} \label{sec:summary}
In this study, we investigate the \hbox{\gam--\mdot} correlation for a sample of \hbox{super-Eddington} accreting quasars, and we compare it to that for a sample of \hbox{sub-Eddington} accreting quasars.
The key points are as following.
\begin{enumerate}
\item
We construct a final sample of 113 \hbox{broad-line}, \hbox{radio-quiet} quasars from the SDSS DR 14 quasar catalog.
We fit their optical spectra to obtain the continuum and the emission-line properties, and we use these properties to estimate the \hbox{black-hole} masses and dimensionless accretion rates.
The X-ray data of our sample are gathered from the \chandra\ and \xmm\ archives, and the \gam\ values are estimated from the \xray\ spectral fitting.
See Section~\ref{sec:s} and~\ref{sec:data}.

\item
We identify a super-Eddington subsample with 38 quasars from our final sample, and we find a statistically significant correlation between \gam\ and \mdot.
We find a significant, but weaker \hbox{\gam--\mdot} correlation for the \hbox{sub-Eddington} subsample that includes 75 quasars.
The correlation slope $(0.34\pm0.11)$ for the \hbox{super-Eddington} subsample is steeper than that $(0.09\pm0.04)$ for the \hbox{sub-Eddington} subsample.
See Section~\ref{sec:gamvsmdot}.

\item
We also find statistically significant correlations between \gam\ and \edd\ for the full sample and \hbox{super-Eddington} super subsample.
The slope of our \hbox{\gam--log\edd} correlation for the full sample is consistent with those of previous studies.
The \hbox{\gam--\edd} correlation for the \hbox{super-Eddington} subsample is stronger than that for the \hbox{sub-Eddington} subsample.
See Section~\ref{sec:gamvsedd}.

\item
We find no apparent correlation between \gam\ and $L_{5100}$.
We find that the correlation between \gam\ and \mbhtd\ is significant for the full sample and the \hbox{sub-Eddington} subsample.
We find no correlation between \gam\ and \mbh\ for the super-Eddington subsample.
See Section~\ref{sec:gamvsmbh}.

\item
Our findings on the \hbox{\gam--\mdot} correlations provide suggestive evidence that the \hbox{disk--corona} connections are different in super- and \hbox{sub-Eddington} accreting quasars.
We propose one qualitative explanation of the steeper \hbox{\gam--\mdot} correlation in the \hbox{super-Eddington} regime that involves larger seed photon fluxes received by the compact coronae from the thick disks in \hbox{super-Eddington} accreting quasars.
See Section~\ref{sec:gammdotdis}.

\end{enumerate}

\subsection{Future Work} \label{sec:futurework}
Larger statistical samples of quasars are required to extend our current study, so that we can confirm our finding of the different \hbox{\gam--\mdot} correlations in super- and \hbox{sub-Eddington} accreting quasars and also explore the underlying physics.
It is important to select super- and \hbox{sub-Eddington} samples and analyze data in an unbiased and systematic manner.
One possibility is to also include higher-redshift SDSS quasars, if we can control or understand the uncertainties on the estimated \hbox{black-hole} masses derived utilizing \ion{Mg}{2} and \ion{C}{4} emission-line properties; obtaining NIR spectroscopy for these \hbox{high-redshift} quasars would also provide a viable way to derive relatively reliable \hbox{black-hole} masses using the \hb\ emission lines.

Our limited sample size is mainly caused by the lack of sensitive \xray\ coverage for the SDSS quasars, as \chandra\ and \xmm\ only cover a small portion of the whole sky.
The \erosita\ telescope \citep[e.g.,][]{Merloni2012,Merloni2019} has the potential for providing good \xray\ observations for the SDSS quasars.
We estimate the number of SDSS DR14 quasars at $z<0.7$ that will have more than 200 2--10~keV net counts from the \erosita\ 4-years all sky survey.
There are 36\,697 SDSS quasars, and we estimate their expected \xray\ fluxes from the \xray--UV correlation of \cite{steffen2006}, adopting an optical spectral slope of $-0.5$ and an \xray\ photon index of 1.8.
GIven the expected limiting flux ($10^{-13}~{\rm erg}~{\rm cm}^{-2}~{\rm s}^{-1}$ in the \hbox{2--10 keV} band) from the \erosita\ 4-years survey \citep{Merloni2012}, approximately 675 quasars will be sufficiently bright to be detected by \erosita\ with more than 200 net counts.
This will substantially increase the sample size for our study presented here.

In the near future, optical spectroscopic surveys such as the SDSS-V\footnote{ https://www.sdss.org/future/.} \citep{Kollmeier2017} and the Dark Energy Spectroscopic Instrument (DESI)\footnote{https://www.desi.lbl.gov.} surveys will provide much larger samples of quasars with optical spectra.
Combining these with the \chandra\ and \xmm\ archives and the \erosita\ data, we will be able to constrain the \hbox{disk--corona} connection in \hbox{super-Eddington} accreting AGNs with greater certainty.

~\\

We thank Qiusheng Gu, Songlin Li, Hezhen Liu and Junjie Mao for helpful discussions.
We thank the referee for providing helpful comments.
We acknowledge financial support from the National Natural Science Foundation of China grants 11991053 and 11673010 (J.H., B.L.), National Key R\&D program of China grant 2016YFA0400702 (J.H., B.L.), and National Thousand Young Talents program of China (B.L.).
\bibliographystyle{aasjournal}
\bibliography{ms_v2}
\clearpage

\begin{deluxetable*}{lcrccccccc}
\tabletypesize{\scriptsize}
\tablewidth{0pt}
\tablecaption{Results of Spearman Rank Correlation Tests and Linear Regression Analyses
}
\tablehead{
\colhead{Relation}   &
\colhead{Sample}   &
\colhead{$R_{\rm S}$}    &
\colhead{$p$}   &
\colhead{$S$}    &
\colhead{$C$}    &  \\
\colhead{(1)}          &
\colhead{(2)}          &
\colhead{(3)}          &
\colhead{(4)}          &
\colhead{(5)}          &
\colhead{(6)}             
}
\startdata

$ {\rm \Gamma}~{\rm vs.}~\dot{\mathscr{M}}                       $&Super-Eddington &  $0.43$  &$  7.75{\times}{ 10^{-3}}  $&$  0.34\pm0.11  $&$  1.71\pm0.17  $ \\
$                                                              $& Sub-Eddington   &  $0.30$  &$  9.98{\times}{ 10^{-3}}  $&$  0.09\pm0.04  $&$  1.93\pm0.04  $ \\
$      $& Full  &  $0.56$  &$  1.10{\times}{\rm 10^{-10}}  $&$  0.13\pm0.02  $&$  1.97\pm0.02  $\\

$  {\rm \Gamma}~{\rm vs.}~\dot{\mathscr{M}}_{\rm NT}                      $&Super-Eddington   &  $0.32$  &$  4.75{\times}{ 10^{-2}}  $&$  0.44\pm0.25  $&$  1.81\pm0.30  $\\
$                                                              $& Sub-Eddington   &  $0.23$  &$  4.78{\times}{ 10^{-2}}  $&$  0.07\pm0.05  $&$  1.93\pm0.04  $ \\
$      $& Full   &  $0.52$  &$  2.69{\times}{\rm 10^{-9}}  $&$  0.15\pm0.02  $&$  2.00\pm0.02  $\\

${\rm \Gamma}~{\rm vs.}~{\lambda_{\rm Edd}}  $& Super-Eddington  &  $0.56$  &$  2.76{\times}{10^{-4}}  $&$  0.59\pm0.16  $&$  2.13\pm0.03  $\\
$                                                              $& Sub-Eddington   &  $0.21$  &$  7.10{\times}{ 10^{-2}}  $&$  0.12\pm0.07  $&$  2.03\pm0.09  $ \\
$               $& Full   &  $0.53$  &$ 1.32{\times}{\rm 10^{-9}}  $&$  0.23\pm0.03  $&$  2.13\pm0.03  $\\

$     {\rm \Gamma}~{\rm vs.}~{\lambda_{\rm Edd,NT}}               $& Super-Eddington   &  $0.49$  &$  1.92{\times}{ 10^{-3}}  $&$  0.73\pm0.27  $&$  2.34\pm0.09  $ \\
$                                                              $& Sub-Eddington   &  $0.15$  &$  2.07{\times}{ 10^{-1}}  $&$  0.09\pm0.08  $&$  1.97\pm0.09  $ \\
$ $& Full    &  $0.50$  &$  2.49{\times}{ 10^{-8}}  $ &$  0.26\pm0.04  $&$  2.18\pm0.04  $\\

$  {\rm \Gamma}~{\rm vs.}~{{M}_{\rm BH}}           $&Super-Eddington   &  $-0.25$  &$  1.35{\times}{10^{-1}}  $ & & \\
$                                                              $&  sub-Edd     &  $-0.30$  &$  6.92{\times}{10^{-3}}  $& &  \\
$      $&  Full     &  $-0.53$  &$  1.20{\times}{\rm 10^{-9}}  $& & \\

$  {\rm \Gamma}~{\rm vs.}~{{M}_{\rm BH,NT}}          $&Super-Eddington   &  $-0.17$  &$  4.16{\times}{10^{-1}}  $& & \\
$                                                              $& \hbox{sub-Eddington}    &  $-0.20$ &$  8.62{\times}{10^{-2}}  $& & \\
$       $&  Full     &  $-0.43$  &$  2.67{\times}{\rm 10^{-6}}  $& &

\enddata

\tablecomments{
Column (1): The parameter used for testing the correlation with \gam; \mdot\ (\newedd) is derived from the mass estimated using the Equation \ref{2}, while \mdotold\ (\tnedd) is derived from the mass estimated using Equation \ref{3}; Column (2): The subsamples and full sample; Column (3): The Spearman rank coefficient; Column (4): The null hypothesis probability; Column (5): The slope of best-fit relation; Column (6): The constant of best-fit relation.
}
\label{tbl-srlr}
\end{deluxetable*}

\end{document}